\documentclass{ieee-ojvt}
\usepackage{amssymb}
\usepackage{cite}
\usepackage{amsmath,amsfonts}
\usepackage{algorithmic}
\usepackage{graphicx,color}
\usepackage{textcomp}
\usepackage{amsthm}
\usepackage{arydshln}
\usepackage{bbding}
\usepackage{bigints}
\usepackage{booktabs}
\usepackage{diagbox}
\usepackage{epsfig,latexsym}
\usepackage{epstopdf}
\usepackage{algorithm}
\usepackage{float}
\usepackage{flushend}
\usepackage{indentfirst}
\usepackage{lastpage}
\usepackage{makecell}
\usepackage{mathtools}
\usepackage{pifont}
\usepackage{psfrag}
\usepackage{setspace}
\usepackage{stfloats}
\usepackage{subfloat}
\usepackage{float}
\usepackage{times}
\usepackage{verbatim}
\usepackage{hyperref}
\usepackage{subfigure}

\theoremstyle{remark}
\newtheorem{theorem}{\quad \textbf{Theorem}}
\newtheorem{lemma}{\quad \textbf{Lemma}}

\newtheorem{remark}{\quad \textbf{Remark}}

\newtheorem{proposition}{\quad \textbf{Proposition}}

\allowdisplaybreaks[4]

\begin{document}
% \receiveddate{XX Month, XXXX}
%\reviseddate{24 April, 2026}
%\accepteddate{XX Month, XXXX}
% \publisheddate{XX Month, XXXX}
% \currentdate{9 December, 2024}
%\doiinfo{OJVT.2024.XXXXX}

\title{Space-Air-Ground-Integrated Networks: The BER vs. Residual Delay and Doppler Analysis}

\author{CHAO~ZHANG\authorrefmark{1} (Member, IEEE),
KUNLUN LI\authorrefmark{1} (Member, IEEE), \\
CHAO XU\authorrefmark{1} (Senior Member, IEEE),
LIE-LIANG YANG\authorrefmark{1} (Fellow, IEEE),\\
AND LAJOS HANZO\authorrefmark{1} (Life Fellow, IEEE)}
\affil{School of Electronics and Computer Science, University of Southampton, Southampton SO17 1BJ, U.K.}
\corresp{CORRESPONDING AUTHOR: Lajos Hanzo (e-mail: lh@ecs.soton.ac.uk).  }

\markboth{Preparation of Papers for IEEE OPEN JOURNAL of VEHICULAR TECHNOLOGY}{C. ZHANG \textit{et al.}}

\begin{abstract}
Perfect Doppler compensation and synchronization is nontrivial due to multi-path Doppler effects and Einstein's theory of relativity in the space-air-ground-integrated networks (SAGINs). Hence, by considering the residual Doppler and the synchronization delay, this paper investigates the bit-error-rate (BER) performance attained under time-varying correlated Shadowed-Rician SAGIN channels. First, a practical SAGIN model is harnessed, encompassing correlated Shadowed-Rician channels, the Snell's law-based path loss, atmospheric absorption, the line-of-sight Doppler compensation, elliptical satellite orbits, and Einstein's theory of relativity. Then, a specific correlation coefficient between the pilot and data symbols is derived in the context of correlated Shadowed-Rician Channels. By exploiting this correlation coefficient, the channel distribution is mimicked by a bi-variate Gamma distribution. Then, a closed-form BER formula is derived under employing least-square channel estimation and equalization for 16-QAM. Our analytical results indicate for a 300-km-altitude LEO that 1) the period of realistic elliptical orbits is around 0.8 seconds longer than that of the idealized circular orbits; and 2) the relativistic delay is lower than 1 $\mu s$ over a full LEO pass (from rise to set). Our numerical results for the L bands quantify the effects of: 1) the residual Doppler; 2) atmospheric shadowing; 3) synchronization errors; and 4) pilot overhead.
\end{abstract}


\maketitle

\section{Introduction}

\begin{figure*}[t]
\centering
\includegraphics[width= 7in]{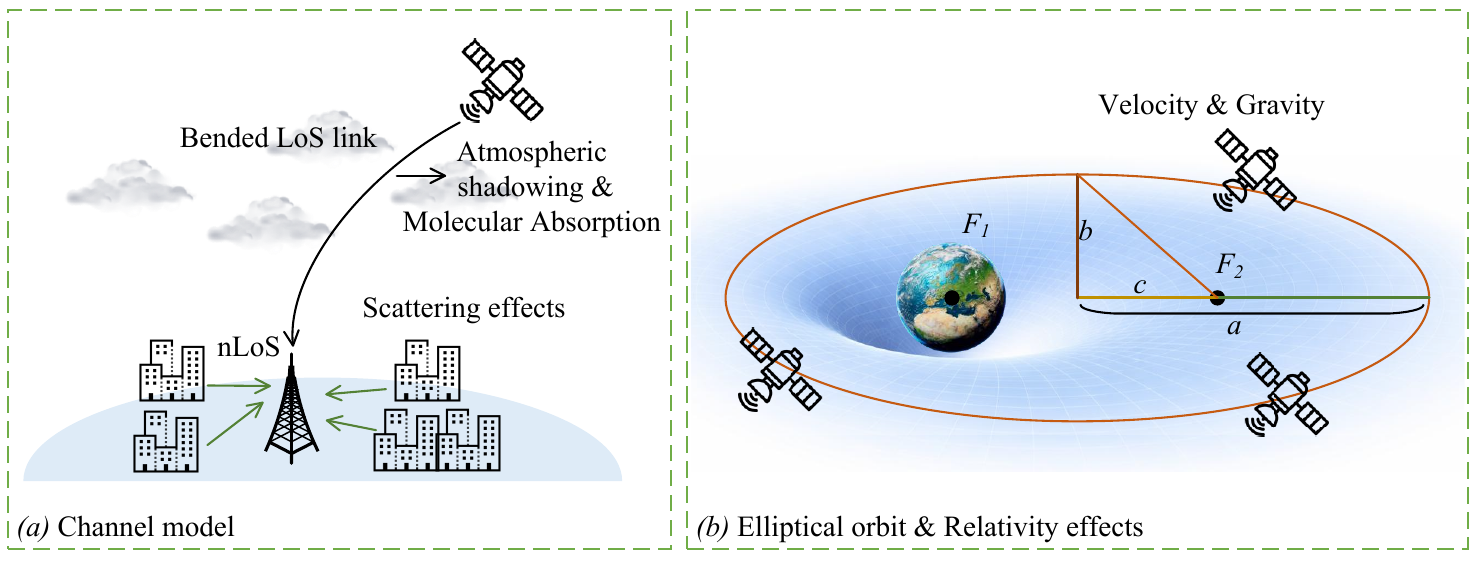}
\caption{Illustration of the system model: (a) Channel model and (b) Elliptical orbits \& Einstein's theory of relativity.}
\label{systemmodel}
\end{figure*}

Presented by the 2025 mobile test summit of Rohde \& Schwarz on ``Satcom 2.0: Navigating the NTN Landscape'', terrestrial communication has covered more than 60\% of the global landmass, but around 40\% still lacks coverage in remote areas, requiring the deployment of enhanced next-generation (NG) solutions \cite{rohde_schwarz_5GAdv_NTN}. Given that the entire landmass accounts for approximately 29\% of the Earth's total surface area, the remaining 71\%, such as the lakes and oceans, poses a significant coverage challenge due to the geographical limitations of terrestrial communication. As popularized by both the academia~\cite{Li_Holographic_2025,Pan_Latency_2023_Sep} and industry (such as Starlink)~\cite{Humphreys_Signal_2023_Oct}, the space-air-ground integrated network (SAGIN) having a dense low-Earth-orbit satellite (LEO) constellation is potentially capable of achieving global coverage, which is likely to be one of the key targets for NG systems~\cite{Huang_Airplane_2019_Sep,Pan_Space_2023_Apr}. Moreover, SAGINs inherently support the emerging vehicle-centric platforms, such as unmanned aerial vehicles (UAVs) and high-altitude platforms (HAPs), which are envisioned as key enablers of NG intelligent transportation systems and for the low-altitude economy. These aerial vehicles require reliable connectivity and accurate sensing capabilities to support applications such as autonomous navigation, real-time monitoring, and dynamic traffic management. By seamlessly integrating the space-, air-, and ground-segments, SAGINs are capable of offering ubiquitous coverage and enhanced service continuity for highly mobile vehicles, which is attractive for supporting heterogeneous transportation systems operating across different altitudes and mobility patterns. However, since the propagation environment of SAGINs is more complex than that of terrestrial communication networks, grave challenges remain to be resolved in order to fulfill the ambitious goals of NG systems.

The first challenge for SAGINs is Doppler compensation. LEOs orbit around the Earth at a velocity of 7.8 km/s approximately on average, resulting in severe Doppler effects. As revealed by state-of-the-art investigations \cite{Zhang_Space_2025_May}, the maximum Doppler shift of a LEO reaches $10^4$ Hz. Furthermore, geosynchronous equatorial orbit satellites (GEOs) have the same angular velocity as that of the Earth's rotation. By contrast, LEOs have much higher velocity and more limited service time (around 5-15 mins for 300-600 km LEOs) for a terrestrial user \cite{Zhang_Space_2025_May}. Hence, maintaining reliable communication under severe Doppler effects within limited service time is one of the main challenges to be tackled. As suggested by 3GPP 38.811~\cite{3GPP_38.811}, since the trajectory of a satellite in a given orbit is predictable, the line-of-sight (LoS) Doppler compensation is capable of eliminating most of the Doppler effects. However, although the LoS Doppler compensation is considered as an effective solution, it is nontrivial to perfectly mitigate the frequency offset due to the multi-path Doppler effects. As a result, the investigation of SAGINs has to migrate from the generally used idealized simplifying assumption of ``perfect Doppler compensation'' to practical investigations in the presence of residual Doppler. In the scenario of isotropic scatterers, it has been validated that the distribution of the maximum residual Doppler obeys Jakes' model \cite{Patzold_modeling_1997_May,Liu_Multi-Scene_2021}. Inspired by the investigation of non-isotropic Doppler scenarios~\cite{Cheng_An_Adaptive_2009_Sep}, the geometry-based stochastic modelling is capable of splitting the non-isotropic Doppler into two components, including the LoS Doppler and residual isotropic Doppler, which echoes the stochastic modelling of the Rician channel. Thus, after the LoS Doppler compensation issues, we exploit Jakes' model to investigate the residual Doppler in SAGINs.

Additionally, although most SAGIN papers assume perfect synchronization, having residual synchronization errors in SAGINs is inevitable \cite{Ashby_Relativity_2003_Jan}. Hence, we reveal in this paper that the circular satellite orbit assumption and purely Newtonian mechanics are insufficient to meet the stringent requirements of precise synchronization. Therefore, elliptical satellite orbits and Einstein's theory of relativity have to be taken into consideration. In this case, the velocity of the satellite in a certain elliptical orbit is not a constant but a function of its position, compared to that in a perfectly circular idealized orbit. Additionally, spatial variations in the gravitational field further aggravate this phenomenon. Hence, the receiver is confronted with continuously time-varying frequency offsets and clock offsets, which may further erode the accuracy of Doppler compensation and clock alignment. As for channel estimation by relying on pilot symbols, given the residual Doppler and time delay~\cite{Cui_Multiobjective_2021_Sep}, the actual system performance is expected to be lower than that attained by harnessing the flat-fading assumption. Hence, the performance evaluation may have to progressively migrate from time-invariant channels to time-varying channels (correlated channels in terms of delay and Doppler) for realistic practical SAGIN scenarios.

Given the popularity of integrated sensing and communications (ISaC), it is timely to highlight the importance of relativistic effects in the context of SAGINs, especially when the delay and Doppler domains are considered \cite{Yang_Doppler_2026}. Unlike conventional communication systems that primarily focus on data transmission, ISaC systems rely critically on precise delay and Doppler estimation for sensing and positioning purposes. In SAGIN scenarios with high-mobility LEO satellites, even small relativity-induced timing and frequency deviations may accumulate over time, leading to non-negligible errors. These accumulated discrepancies directly affect the accuracy of delay-Doppler domain processing, thereby degrading sensing performance in localization and tracking \cite{Ashby_Relativity_2003_Jan}. Therefore, incorporating relativistic effects is essential for developing physically consistent models and for ensuring high-precision ISaC operation in SAGIN environments. This further highlights the necessity of revisiting conventional channel modeling assumptions when moving from communication-centric designs to joint communication and sensing paradigms.

\begin{table*}[thb] \label{table_novelty}
\centering
\caption{THE NOVELTY TABLE}
\begin{tabular}{|p{7.1cm}|c|c|c|c|c|c|c|}
\hline
\textbf{Novelty Points}
& \cite{Abdi_A_New_2003_May}
& \cite{3GPP_38.811}
& \cite{Tiejun_Performance_2026_Jun}
& \cite{Zhang_Space_2025_May}
& \cite{Tang_Effect_1999_Dec,Cao_Closed_2005_Jul,Dong_Symbol_2005_Mar}
& \cite{Bankey_Ergodic_2018_May,Bankey_Performance_2018}
& Our Paper \\
\hline

Shadowed-Rician Channel
& $\checkmark$ &  &  & $\checkmark$ &  & $\checkmark$ & $\checkmark$ \\ \hline

Snell's Law Based Path Loss
&  &  &  & $\checkmark$ &  &  & $\checkmark$ \\ \hline

Atmospheric Absorption
&  &  &  & $\checkmark$ &  &  & $\checkmark$ \\ \hline

LoS Doppler Compensation
&  & $\checkmark$ & $\checkmark$ & $\checkmark$ &  & $\checkmark$ & $\checkmark$ \\ \hline

Residual Doppler Effects (Jakes' Model)
&  &  &  &  & $\checkmark$ &  & $\checkmark$ \\ \hline

M-QAM Modulation (eg. 16-QAM)
&  &  &  &  & $\checkmark$ &  & $\checkmark$ \\ \hline

Time-Varying Channels Due to Doppler Effects
&  &  & $\checkmark$ &  & $\checkmark$ & $\checkmark$ & $\checkmark$ \\ \hline

Imperfect Channel Estimation \& Equalization
&  &  &  &  & $\checkmark$ &  & $\checkmark$ \\ \hline

Correlated Shadowed-Rician Distribution
&  &  &  &  &  & $\checkmark$ & $\checkmark$ \\ \hline \hline

\textbf{Elliptical Orbits \& Circular Orbits}
&  &  &  &  &  &  & $\checkmark$ \\ \hline

\textbf{Einstein's Theory of Relativity}
&  &  &  &  &  &  & $\checkmark$ \\ \hline

\textbf{Closed-Form BER Under the Above Situations }
&  &  &  &  &  &  & $\checkmark$ \\ \hline

\end{tabular}

\end{table*}

\subsection{State-of-the-Art}

\textbf{Doppler Compensation}: Doppler compensation has been widely investigated in the recent decades. Liu~\cite{Liu_Frequency_2003_Mar} has proposed a satellite-referenced synchronization method that effectively eliminates frequency synchronization errors without the employment of guard bands. Zhou et al. \cite{Zhou_A_Simultaneous_2025} have proposed a simultaneous positioning and orbit correction method that jointly estimates the receiver's localization and the satellite orbit's errors to significantly improve localization accuracy. Yeh et al. \cite{Yeh_Efficient_2024_Dec} have proposed a low-complexity multi-stage Doppler shift compensation method for LEO-assisted orthogonal frequency division multiple access (OFDMA) systems. Tanash et al. \cite{Tanash_Statistical_2025} have investigated spherical-geometry-based analytical framework to characterize and statistically model Doppler shift and intra-cell differential Doppler in SAGINs. Since the above papers have assumed perfect synchronization without considering Einstein's theory of relativity, their accuracy may require further analysis. Additionally, the aforementioned algorithms are designed the circular satellite orbits, thus further investigations are required for realistic practical elliptical satellite orbits.

\textbf{Time Synchronization}: As for the clock synchronization, there is a paucity of literature for LEOs. Krieger and Zan \cite{Krieger_Relativistic_2014_Feb} investigated the relativistic time and phase errors caused by the relativity of simultaneity in bistatic and multistatic synthetic aperture radar systems, however the communication performance is beyond the scope of this paper. Additionally, inspired by the recent development of artificial intelligence, deep-learning-based methods may be a promising solution for the complex synchronization tasks in SAGINs \cite{Wang_GaussMask_2025,Liu_Deep_2022_Apr}.

\textbf{Analysis of Correlated Channels}: In the presence of atmospheric effects, the SAGIN channel is modelled by a Shadowed-Rician distribution, thus these correlated Shadowed-Rician channels have to be evaluated \cite{Abdi_A_New_2003_May}. The state-of-the-art investigation of correlated channels has encompassed correlated Rayleigh channels \cite{Tiejun_Performance_2026_Jun,Tang_Effect_1999_Dec}, correlated Rician channels \cite{Cao_Closed_2005_Jul,Dong_Symbol_2005_Mar}, correlated Nakagami-$m$ channels \cite{Shi_Inverse_2017_May}, and correlated Shadowed-Rician channels \cite{Bankey_Ergodic_2018_May,Bankey_Performance_2018}. Since the distribution analysis of correlated Shadowed-Rician channels is very challenging, there are only a few papers investigating the correlated Shadowed-Rician channels. For SAGINs, Bankey and Upadhyay have proposed a Meijer's G-function-based distribution of correlated Shadowed-Rician channels and have investigated the capacity and outage performance \cite{Bankey_Ergodic_2018_May,Bankey_Performance_2018}, however the BER performance analysis in the face of Doppler compensation errors and synchronization delays is still open to a large extent. Additionally, Tang et al. \cite{Tang_Effect_1999_Dec} have proposed a method for evaluating the BER of correlated channels by harnessing the Wiener-Khinchin theorem under isotropic Doppler (Jakes' model), but their evaluation is for terrestrial communications. Hence, by exploiting the Wiener-Khinchin theorem-based methodology, the BER performance under time-vary SAGIN channels has to be evaluated for practical scenarios.

\subsection{Contribution}

In the face of imperfect Doppler compensation and synchronization misalignment, we analyze the theoretical BER under correlated Shadowed-Rician channels in SAGINs. As explicitly indicated by the novelty claimed in Table \ref{table_novelty}, the contributions of this paper are as follows:
\begin{itemize}
\item First, a practical channel model is exploited, including the Earth's rotation, correlated Shadowed-Rician channels, Snell-law-based path loss, and Beer-Lambert-law-based atmospheric absorption. Additionally, a practical elliptical LEO orbit is investigated under Einstein's theory of relativity. LoS Doppler compensation is provided for both the circular and elliptical orbits, followed by the distribution of the residual Doppler, modelled by Jakes' model.
\item Then, the autocorrelation of the Shadowed-Rician channel is evaluated. Since the Shadowed-Rician channel is modelled by the combination of the Nakagami-$m$-faded LoS component and the Rayleigh-faded non-LoS (nLoS) component, we separately derive the autocorrelation of Gaussian variables and that of Nakagami-$m$ variables. By exploiting the autocorrelations, the correlation coefficient of the correlated Shadowed-Rician channel is formulated, which is exploited for formulating its distribution.
\item Additionally, we employ the nearest pilot of the data symbol in order to estimate and to equalize the correlated Shadowed-Rician channel. Hence, the distribution of the pilot symbols and data symbols at the output of the channel are mimicked by a pair of correlated Gamma distributions, generating a bi-variate correlated Gamma distribution. Then, the closed-form BER formulas under time-varying SAGIN channels are derived for 16-quadrature amplitude modulation (16-QAM) under the consideration of residual delay and Doppler effects.
\item For 2GHz carriers, numerical results quantify the effects of: 1) the residual Doppler; 2) atmospheric shadowing; 3) synchronization errors; and 4) pilot overhead.
\end{itemize}

\subsection{Organization}

The paper is organized as follows. Section II provides a practical channel model, encompassing the Snell's law-based path loss, molecular absorption, the correlated Shadowed-Rician distribution, elliptical satellite orbits, and Einstein's relativity. In Section III, we derive the distribution of the correlated Shadowed-Rician channel and our closed-form 16-QAM BER formulas. Numerical results are presented in Section IV, followed by our conclusions in Section V.

\section{The Channel Model For SAGINs}

In Fig. \ref{systemmodel}, we consider elliptical satellite orbits and the relativistic effects. Hence, the SAGIN channel model encompasses: 1) correlated Shadowed-Rician channels; 2) Snell-law-based path loss; 3) Beer-Lambert-law-based atmospheric absorption; 4) LoS Doppler compensation; 5) elliptical orbits; and 6) Einstein's theory of relativity~\cite{Zhang_Space_2025_May}.

\subsection{Small-Scale Fading}

The small-scale fading and atmospheric effects are modelled as a Shadowed-Rician distribution~\cite{Abdi_A_New_2003_May}, formulated as:
\begin{align}\label{complex_signal}
  & {h_{c}}(t) = \underbrace{A(t) \exp \left[ j \zeta (t) \right]}_{nLoS}
+ \underbrace{Z(t) \exp (j \xi)}_{LoS} \notag\\
&  = \left( {{A_I}(t) + Z(t)\sin \xi } \right) + j\left( {{A_Q}(t) + Z(t)\cos \xi } \right),
\end{align}
where the scattering component $A(t)$ is represented by a Rayleigh distribution having uniformly distributed phases $\zeta(t)$. The LoS component $Z(t)$ is modelled as a Nakagami-$m$ distribution with a dominant phase $\xi$. Then, their powers are defined as $\mathbb{E}[A^2(t)] = 2b_0$ and $\mathbb{E}[Z^2(t)] = \Omega$. We additionally have $\mathbb{E}[A_I^2(t)] = \mathbb{E}[A_Q^2(t)]  =b_0$. Thus, the normalized probability density function (PDF) and cumulative distribution function (CDF) of the Shadowed-Rician distribution in the power domain are expressed as:
\begin{align}\label{PDF_h_ST_K}
&{f_{{{\left| {{h_{c} }} \right|}^2}}}(x) =  \sum\limits_{k = 0}^{m - 1} \binom{m-1}{k}\frac{{{{\left( {{K_{Sct}}} \right)}^{m - k - 1}}{{\left( {{K_{LoS}}} \right)}^k}}}{{k!{{\left( {{K_{Sct}} + {K_{LoS}}} \right)}^m}}}\notag\\
&\hspace*{0.3cm} \times {{\left( {\frac{x}{{{K_{Sct}} + {K_{LoS}}}}} \right)}^k}\exp \left( { - \frac{x}{{{K_{Sct}} + {K_{LoS}}}}} \right),\\
\label{CDF_h_ST_K}
&{F_{{{\left| {{h_{c} }} \right|}^2}}}(x) =  1 -  {\sum\limits_{k = 0}^{m - 1} {\binom{m-1}{k}} \frac{{{{\left( {{K_{Sct}}} \right)}^{m - k - 1}}{{\left( {{K_{LoS}}} \right)}^k}}}{{{{\left( {{K_{Sct}} + {K_{LoS}}} \right)}^{m - 1}}}}} \notag\\
& \times \sum\limits_{p = 0}^k {\frac{x^p}{{p!}} {{{\left( {\frac{1}{{{K_{Sct}} + {K_{LoS}}}}} \right)}^p}\exp \left( { - \frac{x}{{{K_{Sct}} + {K_{LoS}}}}} \right)} },
\end{align}
where the shape parameter of the Nakagami-$m$ distribution is assumed to be an integer, denoted as $m$. We also have ${K_{Sct}} = 2{b_0} $ and ${K_{LoS}} = \frac{\Omega }{m}$, where $2 b_0$ is the received power of the nLoS component, and $\Omega$ is the received power of the LoS component. Hence, the normalization is mathematically formulated as $2b_0+\Omega = 1$.

\subsection{Large-Scale Fading}

Based on Snell's Law, the wave propagation under the impact of atmospheric refractivity is considered. Given the refractivity parameters, including ${N_0^{'}} = 315 \times 10^{-6}$ and $h_0 = 7.5$ km, in the ITU-R recommendation \cite{ITU-R_Refraction}, the atmospheric refractive index is formulated as $ n\left( h \right)=  1 + {N_0^{'}}\exp \left( {-\frac{h}{{{h_0}}}} \right)$. Hence, the curved wave trajectory encountered upon entering the atmosphere is formulated as \cite{Zhang_Space_2025_May}:
\begin{align}\label{d_rf}
%{d_{rf}}   = & \int_0^H \frac{n(h)}{\sqrt{1 - \left( \frac{n_0 \cos(\theta_0)}{n(h) \left(1 + \frac{h}{R}\right)} \right)^2 }} dh ,\notag\\
{d_{rf}}=&  \sum\limits_{i = 1}^Q {\frac{{H{\omega _i}}n\left( {{\kappa _i}} \right)}{2}{{\left( {1 - {{\cos }^2}\left( {\frac{2i-1}{{2Q }}\pi } \right)} \right)}^{ \frac{1}{2}}}} \notag\\
&\times {\left( {1 - {{\left( {\frac{{{n_0}\cos \left( {{\theta _0}} \right)}}{{n\left( {{\kappa _i}} \right)\left( {1 + \frac{{{\kappa _i}}}{R}} \right)}}} \right)}^2}} \right)^{ - \frac{1}{2}}},
\end{align}
where we denote the Earth's radius as $R$, the altitude of the satellite's orbit as $H$, and the initial (detected) elevation angle as $\theta_0$.

Subsequently, the path loss is formulated as:
\begin{align}\label{pathloss}
{{\cal P}_{PL}} = {\left( {\frac{c}{{4 \pi {f_c}}}} \right)^2}d_{rf}^{ - \alpha_{pl} },
\end{align}
where $c$ is the speed of light, ${{f_c}}$ is the frequency of carriers, and $\alpha_{pl}$ is the path loss exponent.

\subsection{Atmospheric Absorption}

As for atmospheric absorption, a simplified Beer-Lambert-law-based model is used for the transmittance \cite{Zhang_Space_2025_May}:
\begin{align} \label{abs}
{{\cal P}_{abs}}\left( \tau_i \right) =  {\exp{\left( -\sum\nolimits_i \tau_i \right)}},
\end{align}
where $\tau_i$ is the optical thickness through the atmosphere~\cite{EODG,ROTHMAN_The_2013} and $i$ represented the type of gases.

\subsection{Elliptical Orbit}

As indicated by the Kepler's laws, the satellites are likely to obey an elliptical orbit instead of a circular one. This subsection considers the parameters of the elliptical orbit in \textbf{Theorem \ref{T_E_orbit}}, and then calculates the periodical delay difference between the two types of orbits in \textbf{Remark \ref{R_E_orbit}}.

\begin{theorem} \label{T_E_orbit}
This theorem considers the Newtonian two-body problem (with the constant gravitation $G$), consisting of the Earth's mass $M_E$ and a LEO's mass $m_{LEO}$. We define $\mu = G(M_e+m_{LEO})$ as the gravitational parameter and $\mathbf r(t)$ as the relative position vector between the Earth and the LEO, denoted as $r(t) = \left\lVert\mathbf r(t)\right\rVert$ with its velocity as ${\dot{\mathbf r}(t)}$.

For elliptical orbits, the specific mechanical energy and specific angular momentum are formulated as \cite{Curtis_OrbitalMechanics_2005}:
\begin{align} \label{E-Obit}
\varepsilon  = \frac{1}{2}\left\lVert \dot{\mathbf r} \right\rVert^2 - \frac{\mu}{r},\quad
h  = \left\lVert \mathbf r \times \dot{\mathbf r} \right\rVert,
\end{align}
and the parameters of the elliptical orbit are characterized as:

\begin{enumerate}
  \item The relative motion $\mathbf r(t)$ lies on a Keplerian ellipse in terms of the time. Its semi-major axis $a$, eccentricity $e$, and semi-minor axis $b$ are given by:
        \begin{align}
            a  = -\,\frac{\mu}{2\varepsilon}, \quad
            e  = \sqrt{1 + \frac{2\varepsilon h^2}{\mu^2}}, \quad
            b  = a\sqrt{1-e^2}.
        \end{align}

        The semi-latus rectum $p$ satisfies Eq. (2.43) of \cite{Curtis_OrbitalMechanics_2005} as:
        \begin{equation}
            p = \frac{h^2}{\mu} = a(1-e^2).
        \end{equation}

        We define the true anomaly (denoted by an angle $\theta$) as the angle between the direction of perigee and the LEO's position, centered on the focus of the ellipse. Based on Eq. (2.35) in \cite{Curtis_OrbitalMechanics_2005}, the orbit's equation is formulated as:
        \begin{equation}
            r(\theta) = \frac{p}{1 + e \cos\theta},
            \qquad 0 \le e < 1.
        \end{equation}

  \item Recall that the Earth's physical radius is denoted as $R$. Let us denote the perigee (closest approach) and apogee (farthest distance) altitudes of the satellite above the Earth's surface by $H_p$ and $H_a$, respectively. Then the perigee and apogee radii measured from the Earth's centre are formulated as:
        \begin{equation}
            r_p = R + H_p,\quad r_a = R + H_a.
        \end{equation}

        In terms of $r_p$ and $r_a$, the geometric parameters of the ellipse are formulated as (Example 3.1 in \cite{Curtis_OrbitalMechanics_2005}):
        \begin{align}\label{e_orbit_a}
            a &= \frac{r_p + r_a}{2} = \frac{(R+H_p) + (R+H_a)}{2}, \\
            \label{e_orbit_e}
            e &= \frac{r_a - r_p}{r_a + r_p} = \frac{(R+H_a) - (R+H_p)}{(R+H_a) + (R+H_p)},  \\
            \label{e_orbit_b}
            b &= a\sqrt{1-e^2}.
        \end{align}

  \item The exact circumference $C$ of the elliptical orbit (the trajectory over one revolution) is formulated as:
        \begin{equation}
            C \;=\; 4a\,E(e),
        \end{equation}
        where $E(e)$ is the complete elliptic integral of the second kind (Eq. 17.3.5 in \cite{AbramowitzStegun1965Handbook}) for $0 \le e < 1$:
        \begin{equation}
            E(e) \;=\; \int_{0}^{\pi/2} \sqrt{1 - e^2 \sin^2\varphi}\,d\varphi.
        \end{equation}

  \item Based on (2.73) of \cite{Curtis_OrbitalMechanics_2005}, the orbital period $T$ of a single revolution depends only on $a$ and $\mu$ as given by Kepler's third law:
        \begin{equation}\label{e_orbit_T}
            T = 2\pi \sqrt{\frac{a^3}{\mu}}.
        \end{equation}

  \item Based on (2.71) of \cite{Curtis_OrbitalMechanics_2005}, the instantaneous orbital speed $v$ at distance $r$ from the focus satisfies the \emph{vis--viva} equation:
        \begin{equation}
            v^2 = \mu\left(\frac{2}{r} - \frac{1}{a}\right).
        \end{equation}

        In particular, the speeds at the perigee and apogee are respectively formulated as:
        \begin{align}
            v_p^2 &= \mu \left(\frac{2}{r_p} - \frac{1}{a}\right) = \frac{\mu}{a}\,\frac{1+e}{1-e}, \\
            v_a^2 &= \mu \left(\frac{2}{r_a} - \frac{1}{a}\right) = \frac{\mu}{a}\,\frac{1-e}{1+e},
        \end{align}
        where we have $r_p = a(1-e)$ and $r_a = a(1+e)$.
\end{enumerate}

\begin{proof}
See APPENDIX~A.
\end{proof}
\end{theorem}

\begin{figure}[h]

\centering
\includegraphics[width= 3.5in]{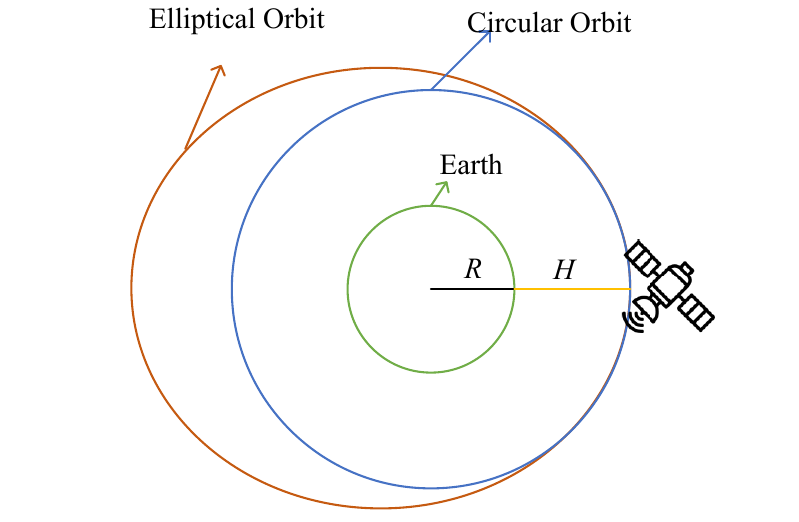}
\caption{The comparison between the circular orbit and elliptical orbit.}
\label{E_Orbit_C_Orbit}

\end{figure}

\begin{remark} \label{R_E_orbit}
In Fig. \ref{E_Orbit_C_Orbit}, we compare the time delay difference between the elliptical orbit and the circular orbit, where we assume that the perigee distance for the elliptical orbit is equal to the radius of the circular orbit. This is formulated as ${r_p} = a\left( {1 - e} \right) = R + H$. Based on \eqref{e_orbit_a} and \eqref{e_orbit_T}, the time delay difference for a period is formulated as:
\begin{align}
 \Delta  T_o &= 2\pi \sqrt {\frac{{{a^3}}}{\mu }}  - 2\pi \sqrt {\frac{{{{\left( {R + H} \right)}^3}}}{\mu }} \notag \\
  & = 2\pi \sqrt {\frac{{{{\left( {R + H} \right)}^3}}}{\mu }} \left( {{{\left( {1 - e} \right)}^{ - \frac{3}{2}}} - 1} \right).
\end{align}

Since the eccentricity of the satellites' orbit is generally calculated in the range of $\left[10^{-4}, 10^{-3}\right]$ \cite{Ruiz_Geosynchronous_2013_Aug,Ruiz_Nearly_2014_Oct}, an example is given to show that the accumulated periodical delay for a single orbit is calculated as:
\begin{equation}
\Delta T_o \in \left[ 0.8135, 8.1440 \right] {\rm{seconds}}, \quad e \in \left[ {{{10}^{ - 4}},{{10}^{ - 3}}} \right],
\end{equation}
where the Earth's radius is $R = 6371$ km, the Earth's mass is $M_E = 5.972\times 10^{24}$ kg, the satellite's mass is $m_{LEO} = 500$ kg, and the gravitational constant is $G = 6.6743 \times 10^{-11} $ $\rm{m^3 kg^{-1} s^{-2}}$.
\end{remark}

\subsection{LoS Doppler Compensation \& Residual Doppler}

The LoS Doppler compensation is investigated for both the circular orbits and elliptical orbits, followed by the distribution of residual Doppler.

\subsubsection{LoS Doppler Compensation for Circular Orbits}

We denote the relative angular velocity between a LEO and a terrestrial user as ${\omega _{R,u}(t)}$ and the maximum elevation angle as $\theta_{max}$, which is obtained at the time $t_0$. With $\psi (t)$ as the central angle subtended after a time $t$ has elapsed, we calculate the angular velocity, formulated as $\dot \psi (t) = \frac{{d\psi (t)}}{{dt}} = {\omega _{R,u}(t)}$. Since the distance from the Earth's center to the LEO is given by $H_{os} =R+H$, the LoS Doppler is formulated as \eqref{LoS_Doppler} with $\psi \left( {t,{t_0}} \right) = \psi (t) - \psi \left( {{t_0}} \right)$ \cite{Zhang_Space_2025_May}.
\begin{figure*}[hpt!]
 \small{
\begin{align}\label{LoS_Doppler}
f_{D,LoS}^{\text{Circ}} =  - \frac{{f_c}}{c}\frac{{R{H_{os}}\sin \left( {\psi \left( {t,{t_0}} \right)} \right)\cos \left( {{{\cos }^{ - 1}}\left( {\frac{{R\cos {\theta _{\max }}}}{{{H_{os}}}}} \right) - {\theta _{\max }}} \right){\omega _{R,u}(t)}}}{{\sqrt {{R^2} + H_{os}^2 - 2R{H_{os}}\cos \left( {\psi \left( {t,{t_0}} \right)} \right)\cos \left( {{{\cos }^{ - 1}}\left( {\frac{{R\cos {\theta _{\max }}}}{{{H_{os}}}}} \right) - {\theta _{\max }}} \right)} }}.
\end{align} }
\hrulefill

\end{figure*}

\subsubsection{LoS Doppler Compensation for Elliptical Orbits}

Again, in elliptical orbits, the velocity of LEOs is not a constant but time-varying. Hence, the LoS Doppler compensation is a function of time $t$, given by \cite{Diniz_An_Algorithm_2024_Oct}:
\begin{equation}
f_{D,LoS}^{\text{Ellip}}(t)
= -\frac{f_c}{c}
\frac{\big(\mathbf v_s(t)-\mathbf v_u(t)\big)\cdot
      \big(\mathbf r_s(t)-\mathbf r_u(t)\big)}
     {\big\|\mathbf r_s(t)-\mathbf r_u(t)\big\|}.
\end{equation}

For a Keplerian elliptical orbit, the radial velocity component is expressed as:
\begin{equation}
v_r(\theta)=\sqrt{\frac{\mu}{a(1-e^2)}}\, e \sin\theta,
\end{equation}
which yields a time-varying Doppler shift proportional to the orbital eccentricity.

\subsubsection{Residual Doppler}

After compensating the LoS Doppler, the main impact of Doppler shift is roughly mitigated~\cite{3GPP_38.811}. Given a LEO at 300-km altitude, the maximum Doppler frequency is at a level of $10^5$ Hz. After LoS Doppler compensation, the residual Doppler may be reduced to hundreds of Hz~\cite{Zhang_Space_2025_May}. We define the residual Doppler as $f_r$ and the maximum Doppler as $f_{r,max}$. By assuming isotropic Doppler due to the random deployment of ground scattering, Jakes' model is employed, whose PDF is formulated as~\cite{Tiejun_Performance_2026_Jun,Haas_Aeronautical_2002_Mar}:
\begin{align}\label{Jakes_Model}
 {f_{{f_r}}}\left( x \right) = \left\{ {\begin{array}{*{20}{c}}
   {\frac{1}{{\pi {f_r}\sqrt {1 - {{\left( {\frac{x}{{{f_r}}}} \right)}^2}} }},} & {\left| x \right| \le {f_{r,max}}} , \\
   {0,} & {{\rm{else}}.}  \\
\end{array}} \right.
\end{align}

\subsection{Einstein's Theory of Relativity}

Einstein's theory of relativity includes the special relativity formula (indicated by \textbf{Proposition \ref{relativity}}) and the general relativity formula (indicated by \textbf{Theorem \ref{relativity2}}).

\begin{proposition}\label{relativity}
\textbf{Special Relativity}: Due to the special relativity, the actual time is slower than the coordinate time measured on Earth. Under circular orbits and a constant velocity $v$, the time difference is given by:
\begin{equation}
\Delta \tau = \Delta t \sqrt{1 - \frac{v^2}{c^2}},
\end{equation}
where $\Delta \tau$ denotes the elapsed time measured by a clock onboard the satellite, $\Delta t$ denotes the elapsed time measured by an observer on Earth, $v$ is the relative velocity of the satellite, and $c$ is the speed of light in a vacuum.
\end{proposition}

\begin{theorem} \label{relativity2}
\textbf{General Relativity}: Upon considering weak-field conditions to keep the $c^{-2}$ items (the Schwarzschild spacetime, formulated in \ref{c-4}), the proper-time increment $d\tau$ and the coordinate-time increment $dt$ satisfy the differential equation \cite{Ashby_Relativity_2003_Jan} formulated as:
\begin{align} \label{relativity_equation}
d\tau  = \frac{{ds}}{c} \simeq dt\sqrt {1 + \underbrace {\frac{{2\left( {V - {\Phi _0}} \right)}}{{{c^2}}}}_{{\rm{Gravity}}} - \underbrace {\frac{{{v^2}}}{{{c^2}}}}_{{\rm{Velocity}}}} ,
\end{align}
where we have:
\begin{align}\label{gravational}
V = -\frac{G M_{\text{E}}}{r} \left[ 1 - J_2 \left( \frac{R}{R+H} \right)^2 P_2(\cos\theta) \right].
\end{align}
The gravitational parameter of the Earth, denoted as $GM_E$ in \eqref{gravational}, is given by: $GM_E = 3.986004418 \times 10^{14} \, \text{m}^3 \text{s}^{-2}$. This value represents the product of the Earth's mass and the Newtonian gravitational constant. Still referring to \eqref{gravational}, the Earth's quadrupole moment coefficient is $J_2 = 1.0826300 \times 10^{-3}$, while the equatorial radius of the Earth is $R = 6.3781370 \times 10^6 \, \text{m}$. The true anomaly $\theta$ is defined as the polar angle measured downward from the axis of rotational symmetry. Finally, $P_2$ of \eqref{gravational} denotes the Legendre polynomial of degree 2, which is explicitly given by: $P_2(x) = \frac{1}{2}(3x^2 - 1)$.

Consequently, the accumulated coordinate-time delay along the clock path is an integral formula, expressed as:
\begin{align}
\int_{{\rm{path}}} {d\tau }  =   \int_{{\rm{path}}} {dt} \sqrt {1 + \frac{{2\left( {V - {\Phi _0}} \right)}}{{{c^2}}} - \frac{{{v^2}}}{{{c^2}}}} .
\end{align}
\begin{proof}
Shown in APPENDIX~B.
\end{proof}
\end{theorem}

\begin{remark} \label{R_E_relativity}
We exploit the realistic visible window calculated in \cite{Zhang_Space_2025_May} for a 300-km-altitude LEO in a circular orbit, expressed as $T = 317$ s. As indicated in \cite{Ashby_Relativity_2003_Jan}, we have ${\Phi _0}/{c^2} = -6.96927 \times 10^{-10}$. Based on \textbf{Proposition \ref{relativity}} and \textbf{Theorem \ref{relativity2}}, we calculated the time delay caused by the relativistic effect between the LEO rising and falling, formulated as:

Special relativity (seconds):
\begin{align}
\tau_s = T- T \sqrt{1 - \frac{v^2}{c^2}} = 1.1  \times 10^{-7}.
\end{align}

General relativity (seconds):
\begin{align}
\tau_s = T- \int_{{\rm{path}}} {dt} \sqrt {1 + \frac{{2\left( {V - {\Phi _0}} \right)}}{{{c^2}}} - \frac{{{v^2}}}{{{c^2}}}}  \approx 3.3  \times 10^{-7},
\end{align}
where we have the velocity of the 300-km-altitude LEO as $v = 7.9035 \times 10^{3}$ m/s, the speed of light as $3\times 10^{8}$ m/s, and the true anomaly as $\theta = \pi/4$.
\end{remark}

\subsection{Received Signal and SNR}

Due to the significant time-varying effects caused by the high Doppler of SAGINs, the number of available pilots within the coherence time is lower than in flat-fading channels. To simplify the derivation, we harness the nearest pilot in terms of time for estimating the channel and equalizing it. Although the noise effect is more significant than upon exploiting multiple pilots, this avoids the performance degradation imposed by the fast variations due to Doppler effects. Hence, the pilot signal received at time $t$ is given by:
 \begin{align}
{y_{s_p}} (t) = {h_{c}}\left( t \right){a_p}{s_p} + n(t),
\end{align}
where ${s_p}$ is the known pilot signal. It is worth noting that we could define ${s_p}=1$ for simplifying the calculation. Additionally, the parameter ${a_p}$ represents the constant power degradation, formulated as ${a_p}=P_s {\mathcal{P}_{PL}}{\mathcal{P}_{abs}}$ with the transmit power $P_s$. The additive white Gaussian noise (AWGN) is denoted as $n(t)$, with ${\sigma ^2}=N_0 B_w$ as its variance, $N_0$ as its power density, and $B_w$ as its effective bandwidth.

The channel coefficient is estimated by pilot symbols at time $t$, formulated as ${{\hat h}_{c}}\left( t \right) = {h_{c}}\left( t \right)$. Hence, to demodulate the signal at the time of $t+\tau$, denoted as $s_s(t+\tau)$, the received and demodulated complex-valued signal is formulated under perfect clock synchronization as:
\begin{align}
 \hat s_s(t+\tau)  = \frac{{{h_{c}}\left( {t + \tau } \right)}}{{{{\hat h}_{c}}\left( t \right)}} s_s(t+\tau) + \frac{n (t+\tau)}{{{{\hat h}_{c}}\left( t \right)}}.
\end{align}

Additionally, as for imperfect synchronization, we define an extra synchronization delay, denoted as $\tau_s$. Hence, the received and demodulated complex-valued signal is formulated as:
\begin{align}
 \hat s_s(t+\tau)  = \frac{{{h_{c}}\left( {t + \tau + \tau_s } \right)}}{{{{\hat h}_{c}}\left( t\right)}} s_s(t+\tau +\tau_s) + \frac{n (t+\tau+\tau_s)}{{{{\hat h}_{c}}\left( t  \right)}}.
\end{align}

The instantaneous receive SNR of the terrestrial user is expressed for a narrow band channel as $\gamma_\text{SNR} = \frac{{a_p} \left| h_c \right|^2 }{\sigma ^2}$. Under the ideal Nyquist sampling, the sampling rate $R_s$ is equal to the bandwidth $B_w$. Hence, we can express the symbol power for modulation as ${E_s} = \frac{{{P_s}}}{{{R_{s}}}} = \frac{{{P_s}}}{B_w}$. For a modulated symbol, the instantaneous SNR is expressed as:
\begin{align}
 {\gamma _{r,\text{symb}}}  = \frac{{{E_s}{{\cal P}_{PL}}{{\cal P}_{abs}}{{\left| h_c \right|}^2}}}{{{N_0}}}
  = \frac{{  {a_p} {{\left| h_c \right|}^2}}}{{{\sigma ^2}}} = \gamma_\text{SNR},
\end{align}
with its average power as $\overline {{\gamma _{r,\text{symb}}}}  = \frac{{a_p}}{{{\sigma ^2}}}\mathbb{E}\left[ {{{\left| h_c \right|}^2}} \right] = \frac{{a_p}}{{{\sigma ^2}}}$.

\section{BER Performance Analysis}

In this section, the autocorrelation of the Shadowed-Rician channel related to the residual Doppler ($f_{r,max}$) and the synchronization delay ($\tau_s$) is evaluated.

Given the tradeoff between analytical tractability and generality, 16-QAM constitutes a popular representative, leading to tractable derivations. Hence, in the following, we derive the closed-form BER formula of 16-QAM. The corresponding parameters in \cite{Hanzo_Adaptive_2002} and the generalization methodology of this paper may be exploited to consider the other modulation schemes.

\subsection{The Autocorrelation of Shadowed-Rician Channels}

Based on the Wiener-Khinchin theorem, the power spectral density of Doppler (Jakes' Model) may be transformed to its autocorrelation in terms of time delay. This is formulated in \textbf{Lemma \ref{L_Gaussian_C}} for Gaussian variables and \textbf{Lemma \ref{L_Gaussian_G}} for Nakagami-$m$ variables, which are employed for deriving the autocorrelation of Shadowed-Rician channels. Then, the correlation coefficient of correlated Shadowed-Rician channels is presented in \textbf{Theorem \ref{T_coefficient}}. By exploiting the correlation coefficient derived, the distribution of correlated Shadowed-Rician channel is mimicked by a bi-variate Gamma distribution with its PDF shown in \textbf{Theorem \ref{T_Gamma_Correlation}}.

\begin{lemma}\label{L_Gaussian_C}
The autocorrelation of Gaussian variables, such as the variables of $A_I(t)$ and $A_Q(t)$, is expressed as~\cite{Tang_Effect_1999_Dec}:
\begin{align}\label{R_A}
{R_{{A_I}}}\left( \tau  \right) = {R_{{A_Q}}}\left( \tau  \right) = {b_0}{\rho _J}\left( \tau  \right),
\end{align}
where the correlation coefficient is ${\rho _J}\left( \tau  \right)= {J_0}\left( {2\pi {f_{r,max}}\tau } \right)$ with $J_0{(\cdot)}$ being the zero-order Bessel function of the first kind and $\tau$ being the time delay.
\begin{proof}
See APPENDIX C.
\end{proof}
\end{lemma}

\begin{lemma}\label{L_Gaussian_G}
A Nakagami-$m$ variable, such as $Z(t)$, can be generated by $2m$ Gaussian variables, formulated as $Z\left( t \right) = \sqrt {\sum\limits_{p = 1}^{2m} {X_p^2} } $, with the $2m$ Gaussian variables denoted as ${X_p} \sim {\cal{N}}\left( {0,\frac{\Omega }{{2m}}} \right)$~\cite{Zhang_Space_2025_May}. The autocorrelation of $Z(t)$ is formulated as:
\begin{align}\label{R_Z}
{R_{Z}}\left( \tau  \right)  =\Omega {\rho _Z}\left( \tau  \right),
\end{align}
where the correlation coefficient is expressed as:
\begin{align}
{\rho _Z}\left( \tau  \right) {=} \frac{1}{m}{\left( {\frac{{\Gamma \left( {m + \frac{1}{2}} \right)}}{{\Gamma (m)}}} \right)^2}{}_2{F_1}\left( { {-} \frac{1}{2}, - \frac{1}{2};m;\rho _J^2\left( \tau  \right)} \right),
\end{align}
and ${}_2{F_1}\left(\cdot,\cdot;\cdot;\cdot\right) $ is the Gaussian hypergeometric function.
\begin{proof}
This lemma is proved in APPENDIX D.
\end{proof}
\end{lemma}

\begin{theorem}\label{T_coefficient}
When considering the residual Doppler effects and synchronization delay, we define the effective time delay between the pilot symbol and data symbol as:
\begin{align}
{\tau _{e}} = \left\{ {\begin{array}{*{20}{c}}
   {\tau ,} & {{\text{Perfect Synchronization}}}  \\
   {\tau  + {\tau _s},} & {{\text{Imperfect Synchronization}}}  \\
\end{array}} \right.  .
\end{align}

Hence, if we have $\alpha  = \left| {{h_{c}}\left( {t + \tau_{e} } \right)} \right|$ and $\hat \alpha =\left| {{{\hat h}_{c}}\left( t \right)} \right| = \left| {{{  h}_{c}}\left( t \right)} \right|$, the correlation between the pair of variables $\alpha^2$ and $\hat \alpha^2$ may be denoted as ${{\mathop{\rm cov}} \left\{ {{\alpha ^2},{{\hat \alpha }^2}} \right\}}$. Based on \textbf{Lemma \ref{L_Gaussian_C}} and \textbf{Lemma \ref{L_Gaussian_G}}, their correlation coefficient is formulated as:
\begin{align}\label{R_Z}
{\rho _{SR}}\left( \tau _{e} \right) = \frac{{{\mathop{\rm cov}} \left\{ {{\alpha ^2},{{\hat \alpha }^2}} \right\}}}{{\sqrt {{\mathop{\rm var}} \left( {{\alpha ^2}} \right){\mathop{\rm var}} \left( {{{\hat \alpha }^2}} \right)} }},
\end{align}
where we have ${{\mathop{\rm cov}} \left\{ {{\alpha ^2},{{\hat \alpha }^2}} \right\}} = 4b_0^2\rho _J^2\left( \tau_{e}  \right) {+} \frac{{{\Omega ^2}}}{m}\rho _J^2\left( \tau_{e}  \right)$ $ {+} 4{b_0}\Omega {\rho _J}\left( \tau_{e} \right){\rho _Z}\left( \tau_{e}  \right)$ and ${\mathop{\rm var}} \left( {{\alpha ^2}} \right)={\mathop{\rm var}} \left( {{{\hat \alpha }^2}} \right) = 4b_0^2 + 6{b_0}\Omega  + \frac{{{\Omega ^2}}}{m}$.
\begin{proof}
Shown in APPENDIX E.
\end{proof}
\end{theorem}

\begin{theorem}\label{T_Gamma_Correlation}
Based on a curve fitting tool, we surmise that $\alpha^2$ and $\hat \alpha^2$ may be modelled as Gamma distributed variables~\cite{Zhang_STAR_2022_Sep}, denoted as $\alpha^2 \sim \Gamma(m_1,\Omega_1)$ and $\hat \alpha^2 \sim \Gamma(m_2,\Omega_2)$. We define ${\gamma _1} = \min \left\{ {{m_1},{m_2}} \right\}$, ${\gamma _2} = \max \left\{ {{m_1},{m_2}} \right\}$, and $\eta  = \rho_{SR}(\tau_{e}) \sqrt {\frac{{{\gamma _2}}}{{{\gamma _1}}}}$. Under a pair of scenarios, including the condition of $m_1=m_2=m$ and that of $m_1  \ne  m_2$, the bi-variate PDFs are respectively represented as:
\begin{align}
\label{PDF_m1_e_m2}
 &f_{  \alpha^2,\hat \alpha^2}({t_1},{t_2};{\gamma _1} = \gamma_2  = \gamma ,\eta )
   \notag \\
  &\hspace*{0.3cm} = \frac{{{{({t_1}{t_2})}^{(\gamma  - 1)/2}}\,{e^{ - \frac{{{t_1} + {t_2}}}{{1 - \eta }}}}}}{{{\eta ^{(\gamma  - 1)/2}}(1 - \eta )\,\Gamma (\gamma )}}  {I_{\gamma  - 1}}\left( {\frac{{2\sqrt {\eta {t_1}{t_2}} }}{{1 - \eta }}} \right),  \\
\label{PDF_m1_ne_m2}
& {f_{{\alpha ^2},{{\hat \alpha }^2}}}\left( {{t_1},{t_2};{\gamma _1},{\gamma _2},\eta } \right)    \notag \\
 &\hspace*{0.3cm} = \frac{{t_1^{{\gamma _1} - \frac{{{\gamma _2}}}{2} - \frac{1}{2}}t_2^{\frac{{{\gamma _2} - 1}}{2}}\exp \left( { - \frac{{{t_1} + {t_2}}}{{1 - \eta }}} \right)}}{{\Gamma \left( {{\gamma _1}} \right)\Gamma \left( {{\gamma _2} - {\gamma _1}} \right)}}\sum\limits_{k = 0}^\infty  {\frac{{{{\left( {1 - \eta } \right)}^{{\gamma _2} - {\gamma _1} - 1}}}}{{{\eta ^{\frac{{{\gamma _2} - k - 1}}{2}}}}}}    \notag  \\
& \hspace*{0.6cm}\times {\left( {\frac{{{t_2}}}{{{t_1}}}} \right)^{\frac{k}{2}}}\frac{{\Gamma \left( {{\gamma _2} {-} {\gamma _1} {+} k} \right)}}{{k!}}{I_{{\gamma _2} {+} k {-} 1}}\left( {\frac{{2\sqrt {\eta {t_1}{t_2}} }}{{1 - \eta }}} \right).
\end{align}

Let us now define a pair of variables, including ${\beta _1} = {\Omega _1}{\gamma _1}$ and ${\beta _2} = {\Omega _2}{\gamma _2}$, to simplify the expression. With the aid of the Jacobian transformation, the bi-variate PDF of $\alpha$ and $\hat \alpha$ is formulated as:
\begin{align}\label{nakagami1}
&f_{\alpha,\hat \alpha}(x,y;\gamma,\eta) \notag\\
&\hspace*{0.3cm}
= 4\beta_1\beta_2 x y\,
f_{\alpha^2,\hat \alpha^2}(\beta_1 x^2,\beta_2 y^2;\gamma,\eta) \notag\\
&\hspace*{0.3cm}
= \frac{4(\beta_1\beta_2)^{(\gamma+1)/2}(xy)^{\gamma}}
{\eta^{(\gamma-1)/2}(1-\eta)\,\Gamma(\gamma)}
\exp\!\left(-\frac{\beta_1 x^2+\beta_2 y^2}{1-\eta}\right) \notag\\
&\hspace*{0.6cm}\times
I_{\gamma-1}\!\left(
\frac{2\sqrt{\eta\beta_1\beta_2}\,x y}{1-\eta}
\right),\qquad x>0,\;y>0. \\
\label{nakagami}
& f_{\alpha,\hat \alpha}(x,y;\gamma_1,\gamma_2,\eta) \notag\\
&\hspace*{0.3cm}
= 4\beta_1\beta_2 x y\,
f_{\alpha^2,\hat \alpha^2}(\beta_1 x^2,\beta_2 y^2;\gamma_1,\gamma_2,\eta) \notag\\
&\hspace*{0.3cm}
= \frac{4\,\beta_1^{\gamma_1-\frac{\gamma_2}{2}+\frac{1}{2}}
\beta_2^{\frac{\gamma_2+1}{2}}
x^{2\gamma_1-\gamma_2}
y^{\gamma_2}}
{\Gamma(\gamma_1)\Gamma(\gamma_2-\gamma_1)}
\exp\!\left(-\frac{\beta_1 x^2+\beta_2 y^2}{1-\eta}\right) \notag\\
&\hspace*{0.6cm}\times
\sum_{k=0}^{\infty}
\frac{(1-\eta)^{\gamma_2-\gamma_1-1}}
{\eta^{\frac{\gamma_2-k-1}{2}}}
\left(\frac{\beta_2}{\beta_1}\right)^{\frac{k}{2}}
\left(\frac{y}{x}\right)^{k}
\frac{\Gamma(\gamma_2-\gamma_1+k)}{k!} \notag\\
&\hspace*{0.4cm}\times
I_{\gamma_2+k-1}\!\left(
\frac{2\sqrt{\eta\beta_1\beta_2}\,x y}{1-\eta}
\right),\qquad x>0,\;y>0.
\end{align}
\begin{proof}
By Eq. (48.23) and (48.24) of \cite{Continuous_Kotz_2000}.
\end{proof}
\end{theorem}

\subsection{The BER of 16-QAM}

In this subsection, \textbf{Lemma \ref{L_16QAM}} presents the conditional BER formula under 16-QAM. Given the residual Doppler effect and the clock offset due to synchronization errors, the average BER under 16-QAM is investigated in \textbf{Theorem \ref{T_BER_MQAM}}.

\begin{lemma}\label{L_16QAM}
Given $\alpha$ and $\hat \alpha$, the conditional BER of 16-QAM is expressed as~\cite{Tang_Effect_1999_Dec}:
\begin{align}\label{BER_16QAM_conditional}
  \mathbb{P}(E\left| {\alpha ,\hat \alpha } \right.)  & = \frac{1}{4}Q\left( {\frac{{3d\alpha }}{{{\sigma _n}}}} \right) + \frac{1}{4}Q\left( {\frac{{d\alpha }}{{{\sigma _n}}}} \right) \notag \\
&\hspace*{-0.5cm} + \frac{1}{4}Q\left( {\frac{{3d\alpha  - 2d\hat \alpha }}{{{\sigma _n}}}} \right)  - \frac{1}{4}Q\left( {\frac{{3d\alpha  + 2d\hat \alpha }}{{{\sigma _n}}}} \right) \notag \\
& \hspace*{-0.5cm}  + \frac{1}{4}Q\left( {\frac{{ - d\alpha  + 2d\hat \alpha }}{{{\sigma _n}}}} \right) + \frac{1}{4}Q\left( {\frac{{d\alpha  + 2d\hat \alpha }}{{{\sigma _n}}}} \right),
\end{align}
where the Gaussian-Q function is expressed as $Q(\cdot)$.
\begin{proof}
See Eq. (24) of \cite{Tang_Effect_1999_Dec}.
\end{proof}
\end{lemma}

\begin{theorem}\label{T_BER_MQAM}
Firstly, given the conditional BER in \textbf{Lemma \ref{L_16QAM}}, the average BER expression is formulated as:
\begin{align}
{\mathbb{P}}_E =  \int_0^{\infty } {\int_0^{\infty }  } \mathbb{P}\left( {E|\alpha ,\hat \alpha } \right){f_{\alpha ,\hat \alpha }}\left( {x,y;{\gamma _1},{\gamma _2},\eta } \right)d\alpha d\hat \alpha.
\end{align}

Upon relying on polar coordinates, we can replace the parameters in \eqref{nakagami}, by considering $ x = \sqrt {\frac{{1 - \eta }}{{{\beta _1}}}} r\cos \theta$ and $y = \sqrt {\frac{{1 - \eta }}{{{\beta _2}}}} r\sin \theta$ with $r \in \left[ {0,\infty } \right]$ and $\theta  \in \left[ {0,\frac{\pi }{2}} \right]$. Then, the closed-form BER expression becomes:
\begin{align}\label{BER_MQAM}
 &{{\mathbb{P}}_E} = \frac{1}{4}\Upsilon \left( {\frac{{3d}}{{{\sigma _n}}},0} \right) {+} \frac{1}{4}\Upsilon \left( {\frac{d}{{{\sigma _n}}},0} \right) {+} \frac{1}{4}\Upsilon \left( {\frac{{3d}}{{{\sigma _n}}}, - \frac{{2d}}{{{\sigma _n}}}} \right) \notag \\
 & {-} \frac{1}{4}\Upsilon \left( {\frac{{3d}}{{{\sigma _n}}},\frac{{2d}}{{{\sigma _n}}}} \right) {+} \frac{1}{4}\Upsilon \left( {\frac{{ - d}}{{{\sigma _n}}},\frac{{2d}}{{{\sigma _n}}}} \right) {+} \frac{1}{4}\Upsilon \left( {\frac{d}{{{\sigma _n}}},\frac{{2d}}{{{\sigma _n}}}} \right),
\end{align}
where we have two types of integrations, expressed as:
 \begin{align}
 \Upsilon \left( {z,0} \right) &= \int_{r = 0}^\infty  {\int_{\theta  = 0}^{\frac{\pi }{2}} {Q\left( {zx} \right)} } f\left( {r,\theta ;{\gamma _1},{\gamma _2},\eta } \right)drd\theta \\
 \Upsilon \left( {z,\zeta} \right) & = \int_{r = 0}^\infty  {\int_{\theta  = 0}^{\frac{\pi }{2}} {Q\left( {zx + \zeta y} \right)} } f\left( {r,\theta ;{\gamma _1},{\gamma _2},\eta } \right)drd\theta  .
 \end{align}
The corresponding closed-form expressions are formulated as \eqref{U1} and \eqref{U2}. In \eqref{U1} and \eqref{U2}, the constituent functions are defined as $g\left( \theta  \right) = \sqrt \eta  \sin \left( {2\theta } \right)$, $f\left( {{\kappa _i},\theta ,z} \right) = 1 + \frac{{{z^2}\left( {1 - \eta } \right){{\cos }^2}\left( \theta  \right)}}{{2{\beta _1}{{\sin }^2}{\kappa _i}}}$, and ${f_2}\left( {{\kappa _i},{\kappa _j},z,\zeta } \right) = \frac{{{{\left( {z\beta _1^{ - 1/2}\cos {\kappa _j} + \zeta \beta _2^{ - 1/2}\sin {\kappa _j}} \right)}^2}\left( {1 - \eta } \right)}}{{2{{\sin }^2}{\kappa _i}}} + 1$. Additionally, the parameters in \eqref{U1} and \eqref{U2} are expressed as: ${t_i} = \cos \left( {\frac{{2i - 1}}{{2Q}}\pi } \right) $, ${\kappa _i} = \frac{\pi }{4}\left( {{t_i} + 1} \right)$, ${t_j} = \cos \left( {\frac{{2j - 1}}{{2Q}}\pi } \right)$, ${\kappa _j} = \frac{\pi }{4}\left( {{t_j} + 1} \right)$, ${w_i}={w_j} = \frac{\pi }{Q}$, $l\left( k \right) = \frac{{\pi {{\left( {1 - \eta } \right)}^{{\gamma _2}}}\Gamma \left( {{\gamma _2} - {\gamma _1} + k} \right)\Gamma \left( {{\gamma _1} + {\gamma _2} + k} \right)}}{{{2^{{\gamma _2} + k + 2}}{\eta ^{\frac{{{\gamma _2} - k - 1}}{2}}}\Gamma \left( {{\gamma _1}} \right)\Gamma \left( {{\gamma _2} - {\gamma _1}} \right)\Gamma \left( {{\gamma _2} + k} \right)k!}}$, and $Q=100$ representing the Chebyshev-Gauss computation times (Section 4.6 of \cite{Press2007Numerical}).

\begin{figure*}[htp!]
\begin{align}\label{U1}
 \Upsilon \left( {z,0} \right) = & \sum\limits_{i = 1}^Q {{w_i}} \sqrt {1 - t_i^2} \sum\limits_{j = 1}^Q {{w_j}} \sqrt {1 - t_j^2} \sum\limits_{k = 0}^\infty  {l\left( k \right)} \frac{{{{\cos }^{2{\gamma _1}}}\left( {{\kappa _j}} \right){{\tan }^{{\gamma _2} + k}}\left( {{\kappa _j}} \right){g^{{\gamma _2} + k - 1}}\left( {{\kappa _j}} \right)}}{{{f^{{\gamma _1} + {\gamma _2} + k}}\left( {{\kappa _i},{\kappa _j},z} \right)}} \notag\\
& \times {}_2{F_1}\left( {\frac{{{\gamma _1} + {\gamma _2} + k}}{2},\frac{{{\gamma _1} + {\gamma _2} + k + 1}}{2},{\gamma _2} + k;\frac{{{g^2}\left( {{\kappa _j}} \right)}}{{{f^2}\left( {{\kappa _i},{\kappa _j},z} \right)}}} \right), \\
\label{U2}
 \Upsilon \left( {z,\zeta } \right) &= \sum\limits_{i = 1}^Q {{w_i}} \sqrt {1 - t_i^2} \sum\limits_{j = 1}^Q {{w_j}} \sqrt {1 - t_j^2} \sum\limits_{k = 0}^\infty  {l\left( k \right)} \frac{{{{\cos }^{2{\gamma _1}}}\left( {{\kappa _j}} \right){{\tan }^{{\gamma _2} + k}}\left( {{\kappa _j}} \right){g^{{\gamma _2} + k - 1}}\left( {{\kappa _j}} \right)}}{{f_2^{{\gamma _1} + {\gamma _2} + k}\left( {{\kappa _i},{\kappa _j},z,\zeta } \right)}}\notag \\
& \times {}_2{F_1}\left( {\frac{{{\gamma _1} + {\gamma _2} + k}}{2},\frac{{{\gamma _1} + {\gamma _2} + k + 1}}{2},{\gamma _2} + k;\frac{{{g^2}\left( {{\kappa _j}} \right)}}{{f_2^2\left( {{\kappa _i},{\kappa _j},z,\zeta } \right)}}} \right).
\end{align}
\hrulefill
\end{figure*}
\begin{proof}
Proved in Appendix~F.
\end{proof}
\end{theorem}

\section{Numerical Results}

In this section, we set the numerical settings as follows. The Earth' radius $R$ is 6371.393 km. The LEO's altitude $H$ is 300 km, with the elevation angle of $\theta_0=80^{\circ}$. The transmit power $P_s$ is within $[18,36]$ dBW. As for the S band, we calculate the noise power, formulated as $\sigma^2 = -170+10\times \log_{10}(BW)=-100$ dBm given by its bandwidth $BW = 10$ MHz for under 2-GHz carriers. We set the path loss component to $\alpha_{pl}=2$, the Nakagami-$m$ channel parameter to $m=8$, and the Rician factor to $K = \Omega/(2b_0)=12$ dB. The maximum residual Doppler is 100 Hz and the effective time delay (owing to synchronization errors) is set to $\tau_{e}=0.2$ ms.

\begin{figure*}[!t]
    \centering

    \subfigure[Simulation: Joint PDF]{
        \includegraphics[width=0.23\textwidth]{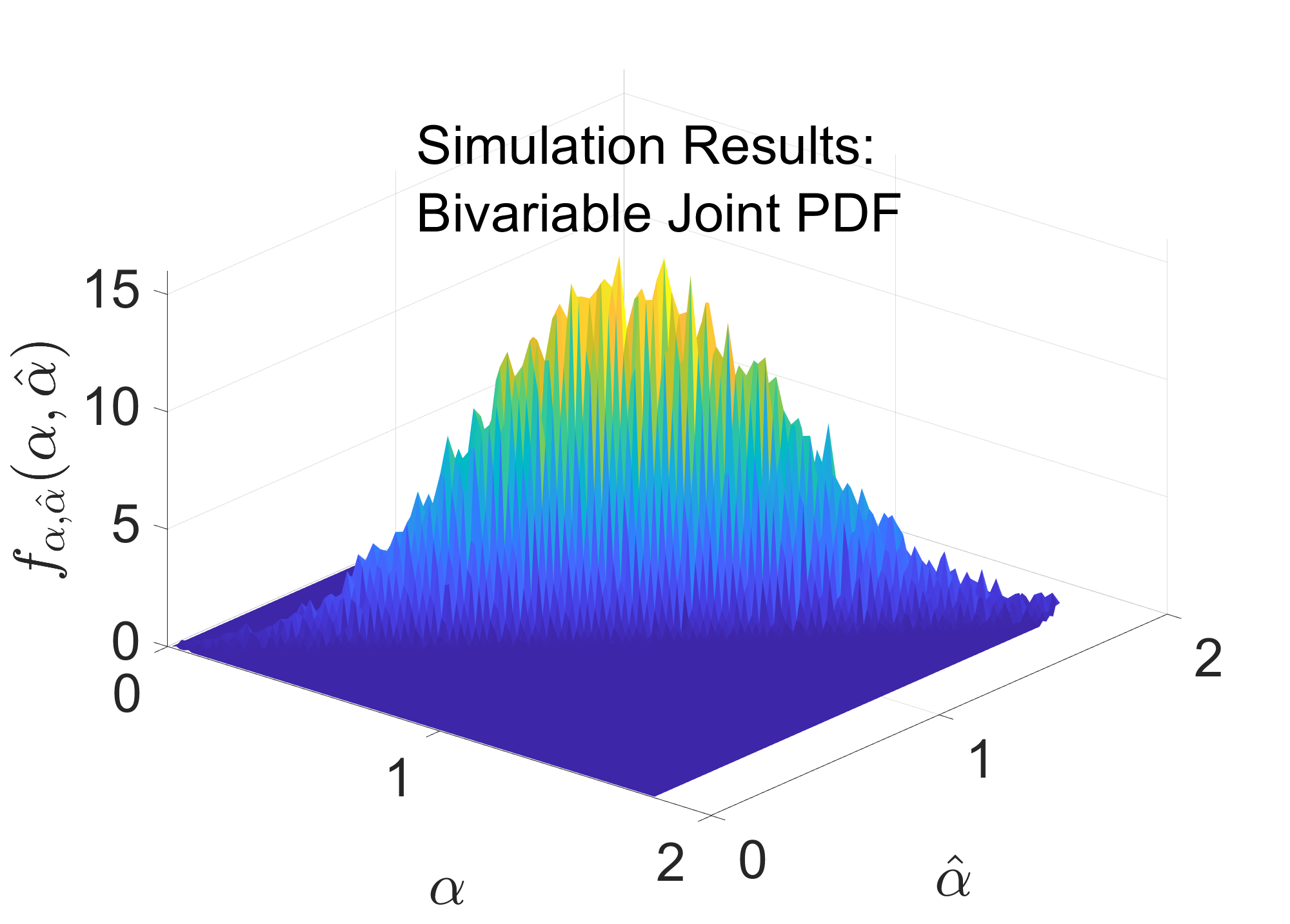}
    }\hfill
    \subfigure[Analysis: Joint PDF]{
        \includegraphics[width=0.23\textwidth]{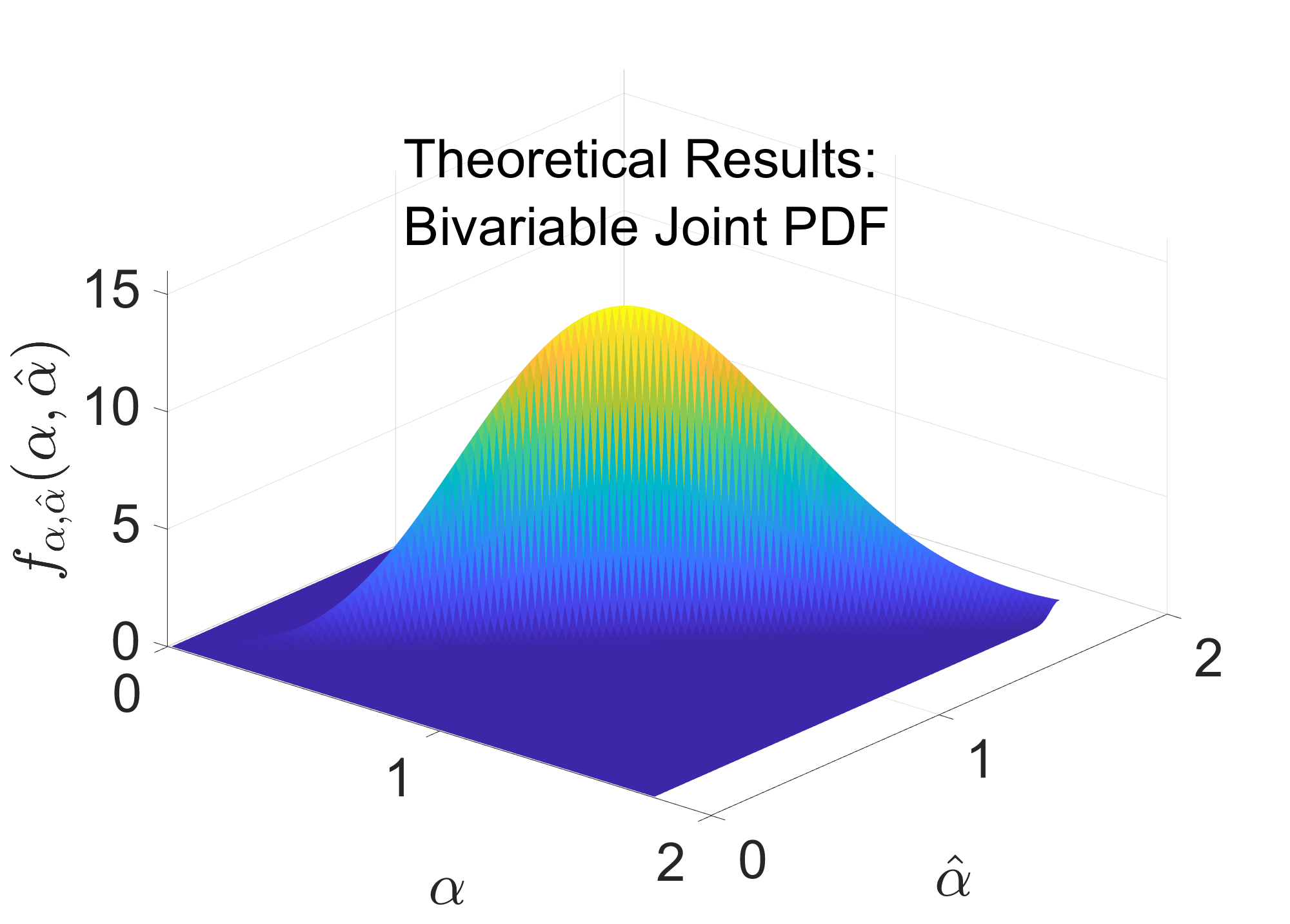}
    }\hfill
    \subfigure[Marginal PDF for $\alpha$]{
        \includegraphics[width=0.23\textwidth]{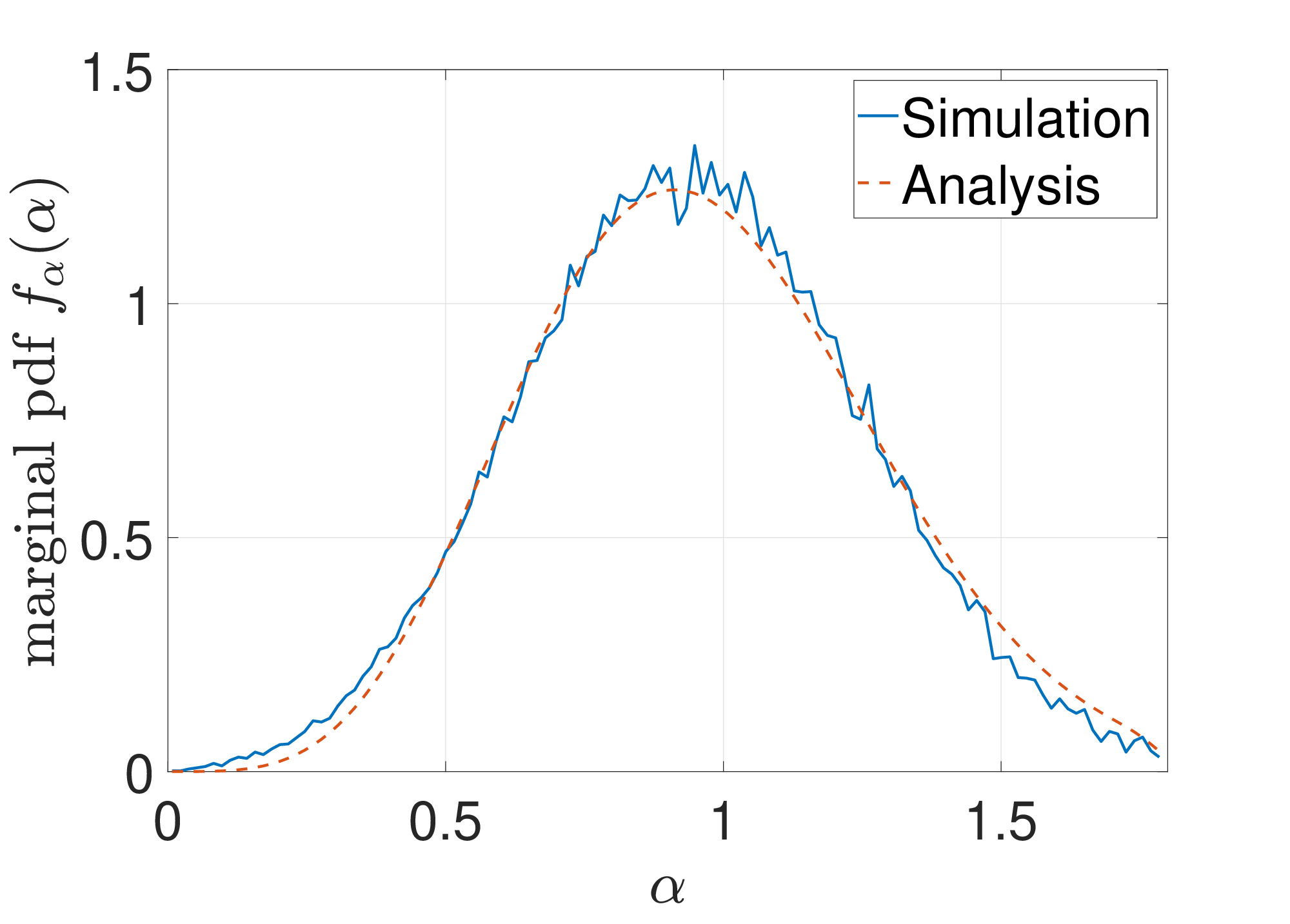}
    }\hfill
    \subfigure[Marginal PDF for $\hat \alpha$]{
        \includegraphics[width=0.23\textwidth]{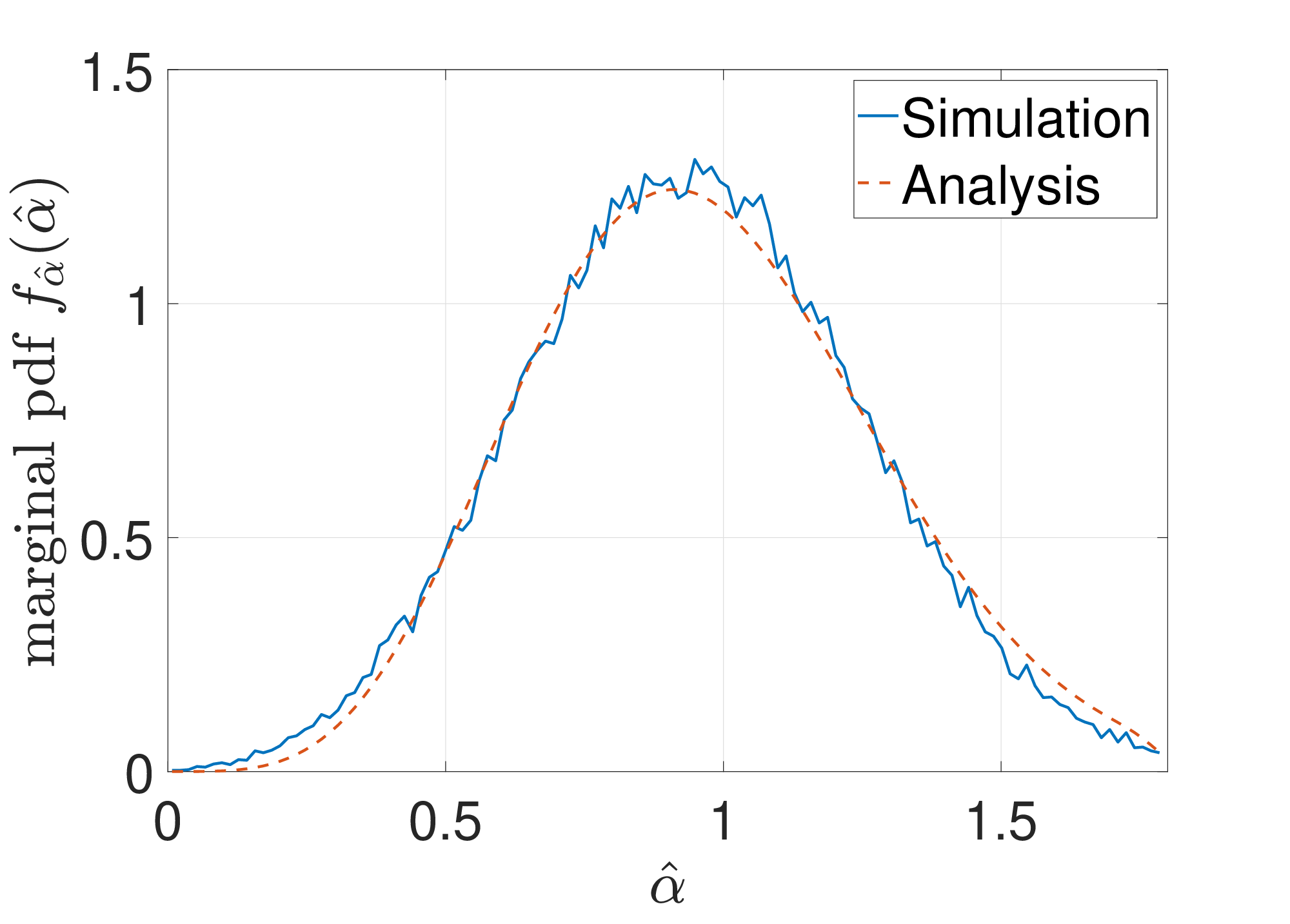}
    }

   \subfigure[Simulation: Joint CDF]{
        \includegraphics[width=0.23\textwidth]{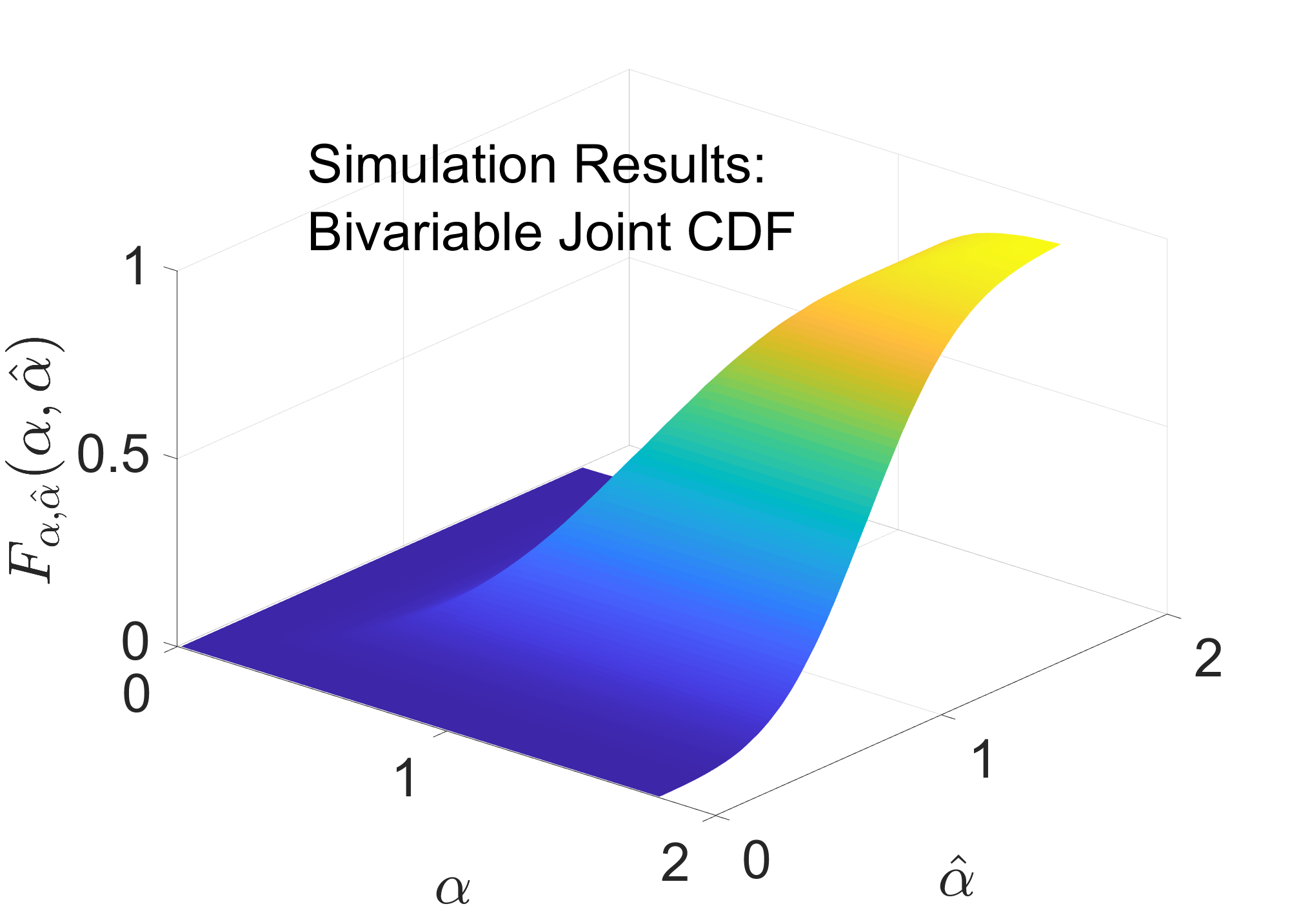}
    }\hfill
    \subfigure[Analysis: Joint CDF]{
        \includegraphics[width=0.23\textwidth]{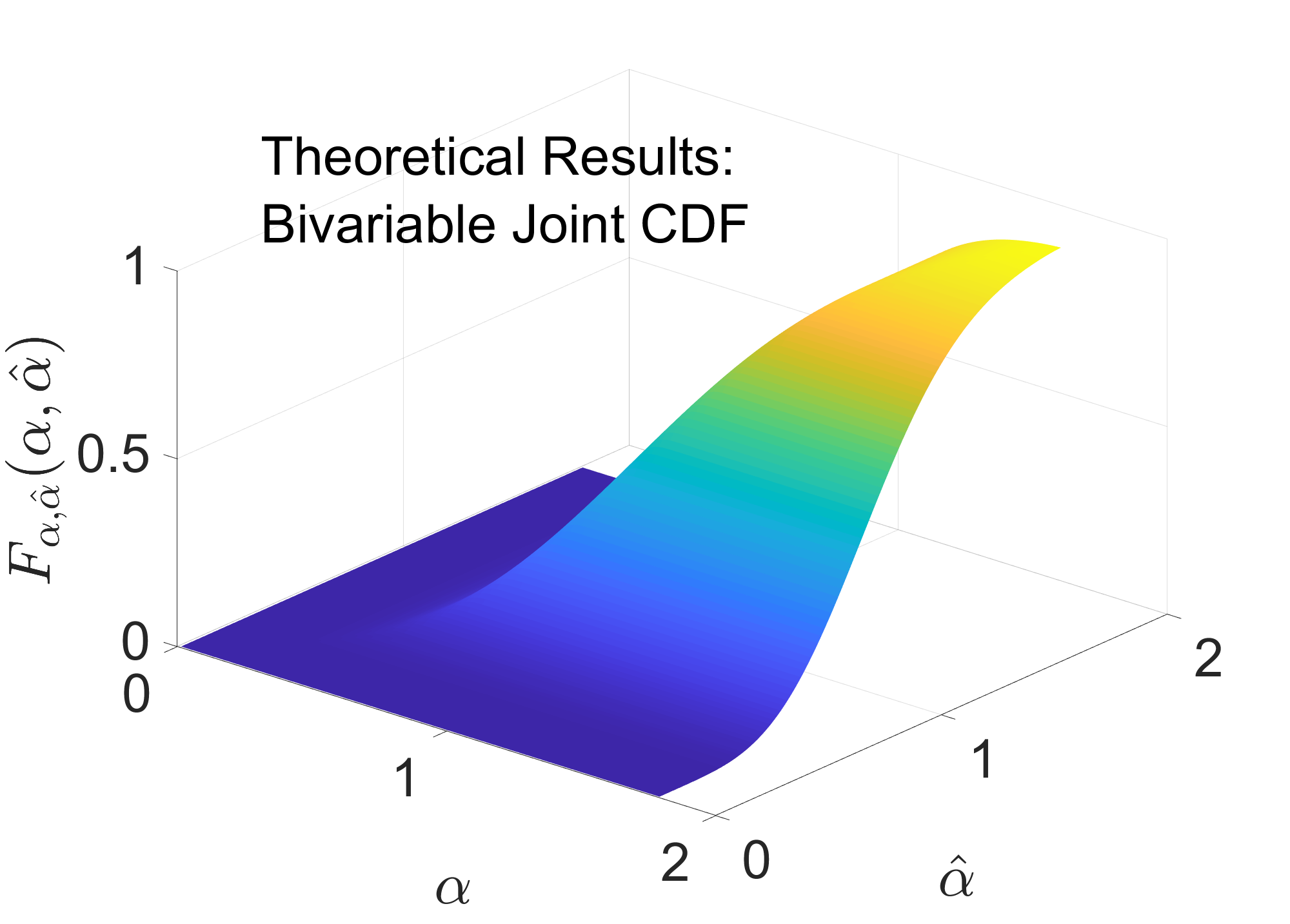}
    }\hfill
    \subfigure[Marginal CDF for $\alpha$]{
        \includegraphics[width=0.23\textwidth]{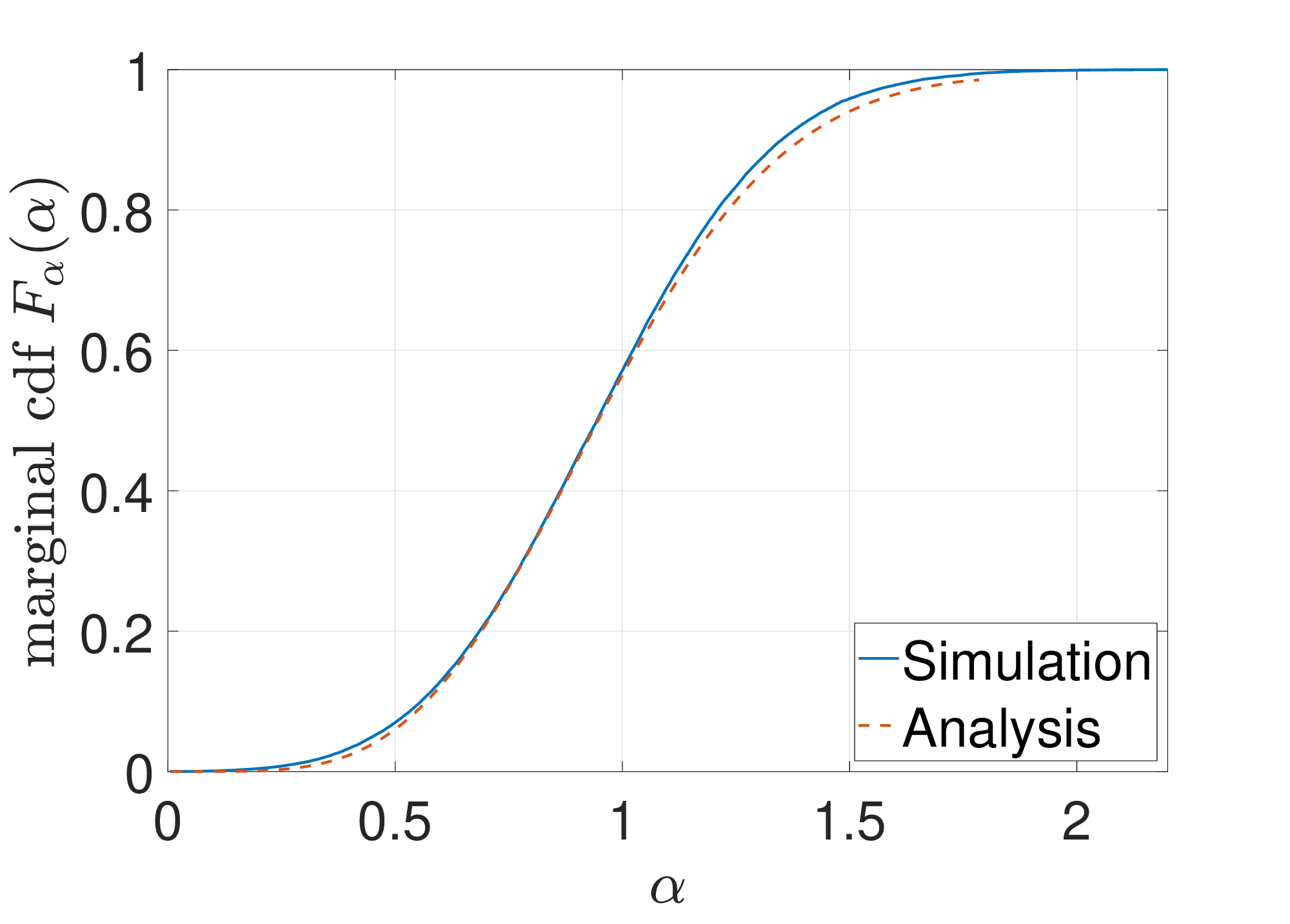}
    }\hfill
    \subfigure[Marginal CDF for $\hat \alpha$]{
        \includegraphics[width=0.23\textwidth]{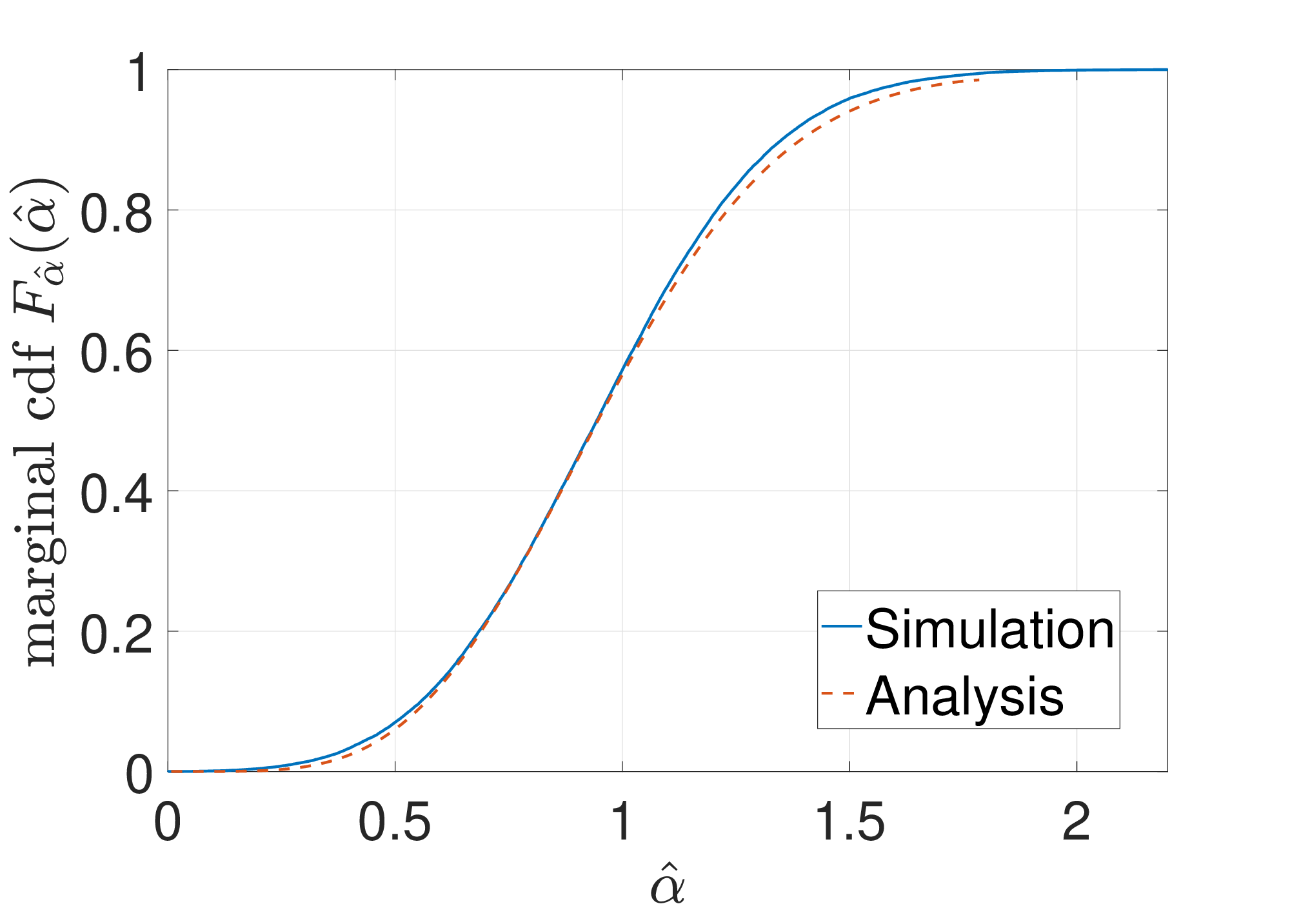}
    }

    \caption{
    The comparison between Monte Carlo simulations and analytical results for the bi-variate PDF and CDF:
    Fig. (a)--(d) are the simulated and theoretical joint PDFs and marginal PDFs;
    Fig. (e)--(h) express the simulated and theoretical joint CDFs and marginal CDFs.
    }
    \label{fig:joint_pdf_cdf}
\end{figure*}

Fig. \ref{fig:joint_pdf_cdf} (with $P_w = 18$ dBW) validates that the distribution of correlated Shadowed-Rician channels may be accurately mimicked by a bi-variate correlated Gamma distribution. In detail, Fig. \ref{fig:joint_pdf_cdf}.(a)-(d) investigates the accuracy of the joint PDF of correlated Shadowed-Rician channels. Fig. \ref{fig:joint_pdf_cdf}.(e)-(h) simulates the CDF of correlated Shadowed-Rician channels. The simulation results show that the bi-variate Gamma distribution obtained by the curve fitting tool closely matches the actual distribution, with only minor discrepancies. Consequently, only small deviations are expected between the analytical results and the simulation results, which are mainly attributed to the curve fitting in accuracy.

\begin{figure}[!htb]
\centering
      \includegraphics[width= 3.5 in]{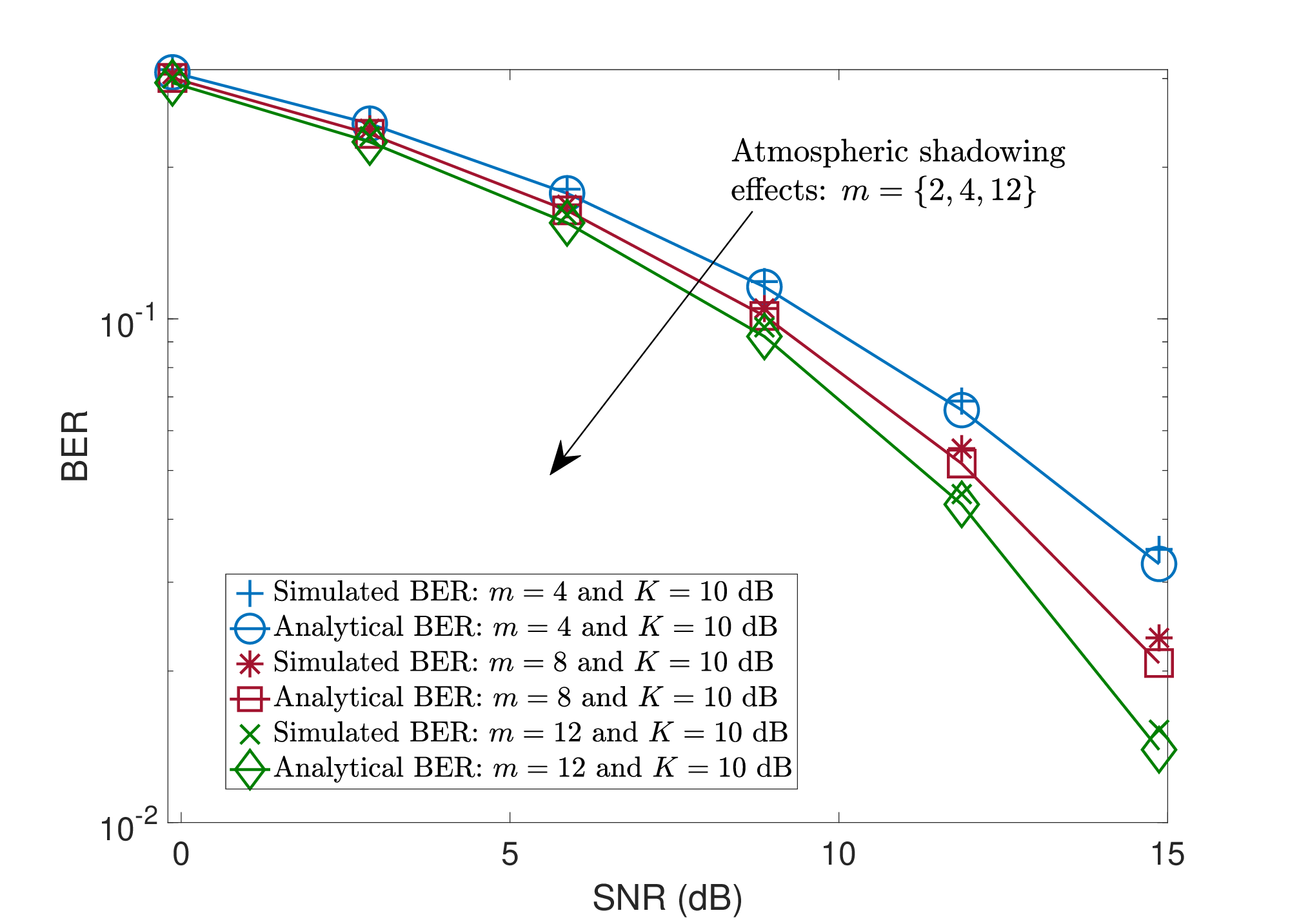}
    \caption{The BER performance versus the receive SNR with different atmospheric shadowing effects ($\bf{m= \{2,4,12\}}$). (\textbf{Theorem \ref{T_Gamma_Correlation}} and \textbf{Theorem \ref{T_BER_MQAM}})}
        \label{fig:BER_SNR_m}
\end{figure}

\begin{figure}[!htb]
\centering
    \includegraphics[width= 3.5  in]{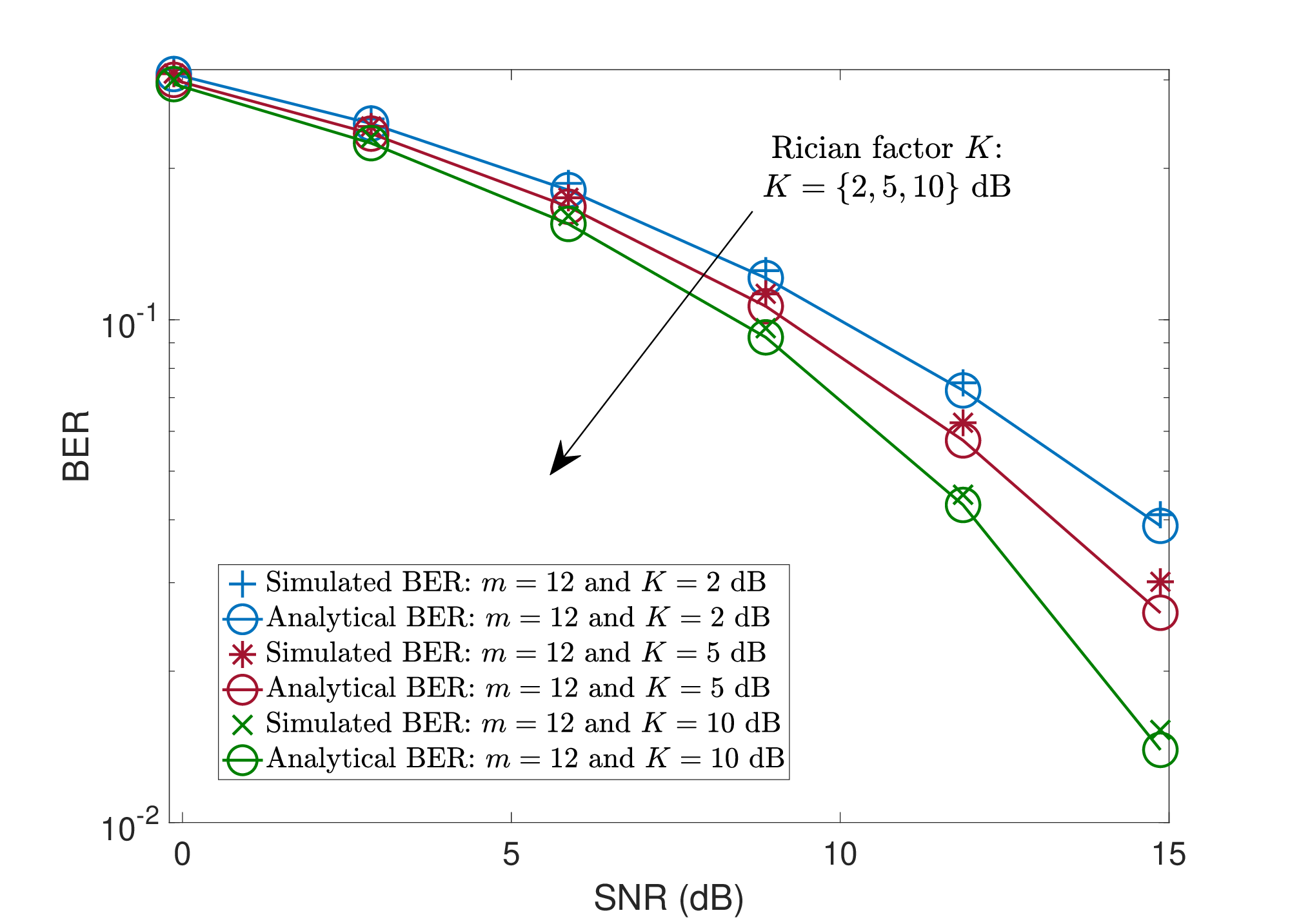}
    \caption{The BER performance versus the receive SNR with the different Rician factor ($\bf{K= \{2,5,10\}}$ dB). (\textbf{Theorem \ref{T_Gamma_Correlation}} and \textbf{Theorem \ref{T_BER_MQAM}})}
        \label{fig:BER_SNR_K}
\end{figure}

In Fig. \ref{fig:BER_SNR_m}, we analyze the BER performance versus the receive SNR, exploiting the settings of $P_w \in \left[ 18, 33\right]$ dBW and $m\in \{2,4,12\}$. Under the consideration of the residual Doppler and delay, the BER after the channel estimation and equalization processes is still above $10^{-2}$ for a receive SNR below 15 dB. Naturally, when the Nakagami-$m$ parameter $m$ increases, the BER performance improves because a higher $m$ indicates stronger LoS component with lower atmospheric shadowing effects. Fig. \ref{fig:BER_SNR_K} investigates the BER performance versus the receive SNR for different values of the Rician factor $K \in \{2,5,10\}$. As expected, increasing the Rician factor $K$ improves the BER because the LoS component is increased observably comparing the results of Fig. \ref{fig:BER_SNR_m} and Fig. \ref{fig:BER_SNR_K}, so that large values of $m$ and $K$ lead to the best BER performance, which represents the practical scenario along with dominant LoS component and weak atmospheric effects.

\begin{figure}[!htb]
\centering
     \includegraphics[width= 3.5 in]{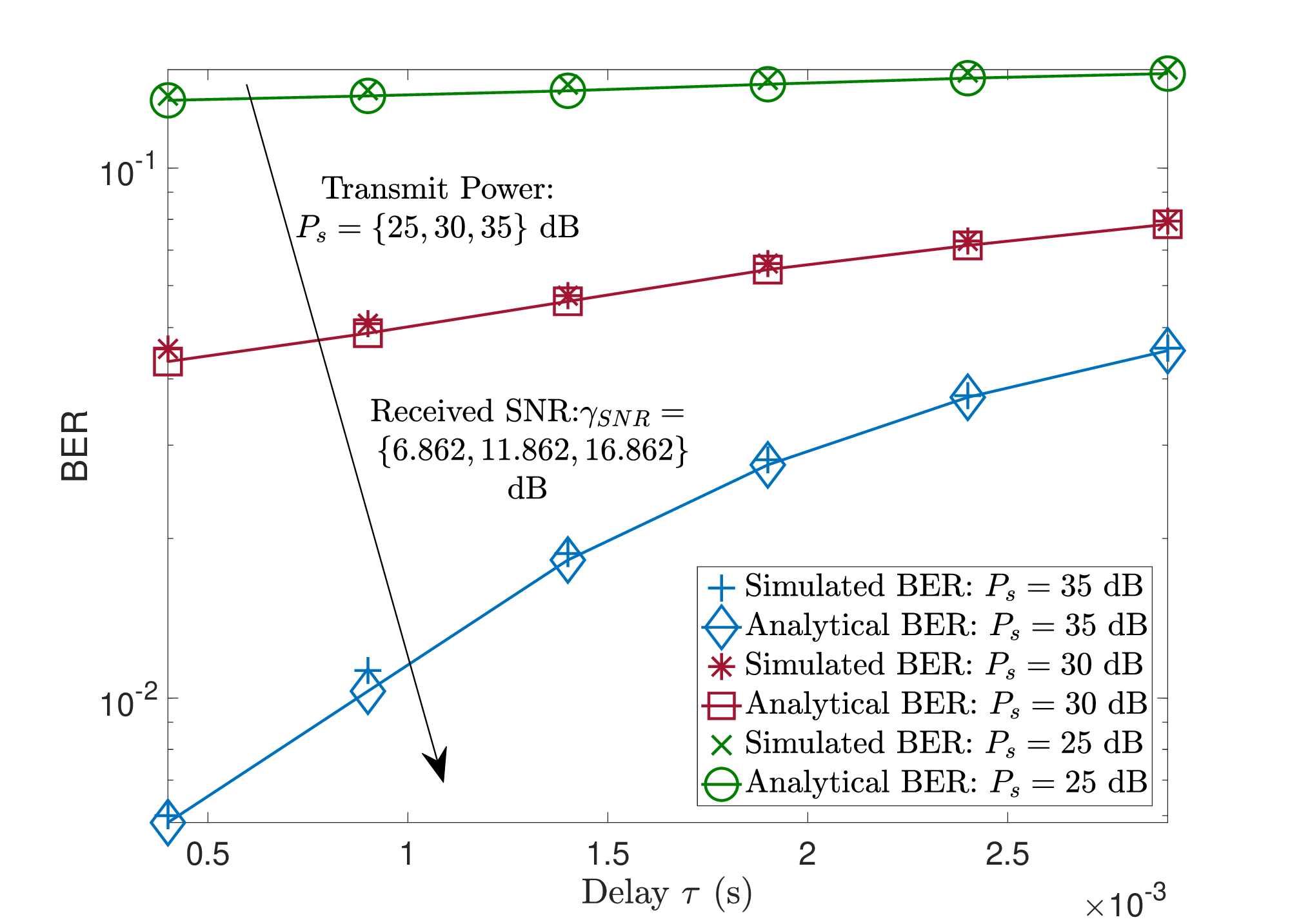}
    \caption{The BER performance versus the delay $\bf{\tau_{e} \in \left[0.4,3.1\right]}$ ms) between the pilot symbol and the signal symbol with the different transmit power (different receive SNR), denoted as $\bf{P_s= \{25,30,35\}}$ dB ($\bf{\gamma_{SNR}= \{6.820,11.820,16.820\}}$ dB). }
        \label{fig:BER_tau_Ps}
\end{figure}

\begin{figure}[!htb]
\centering
    \includegraphics[width= 3.5 in]{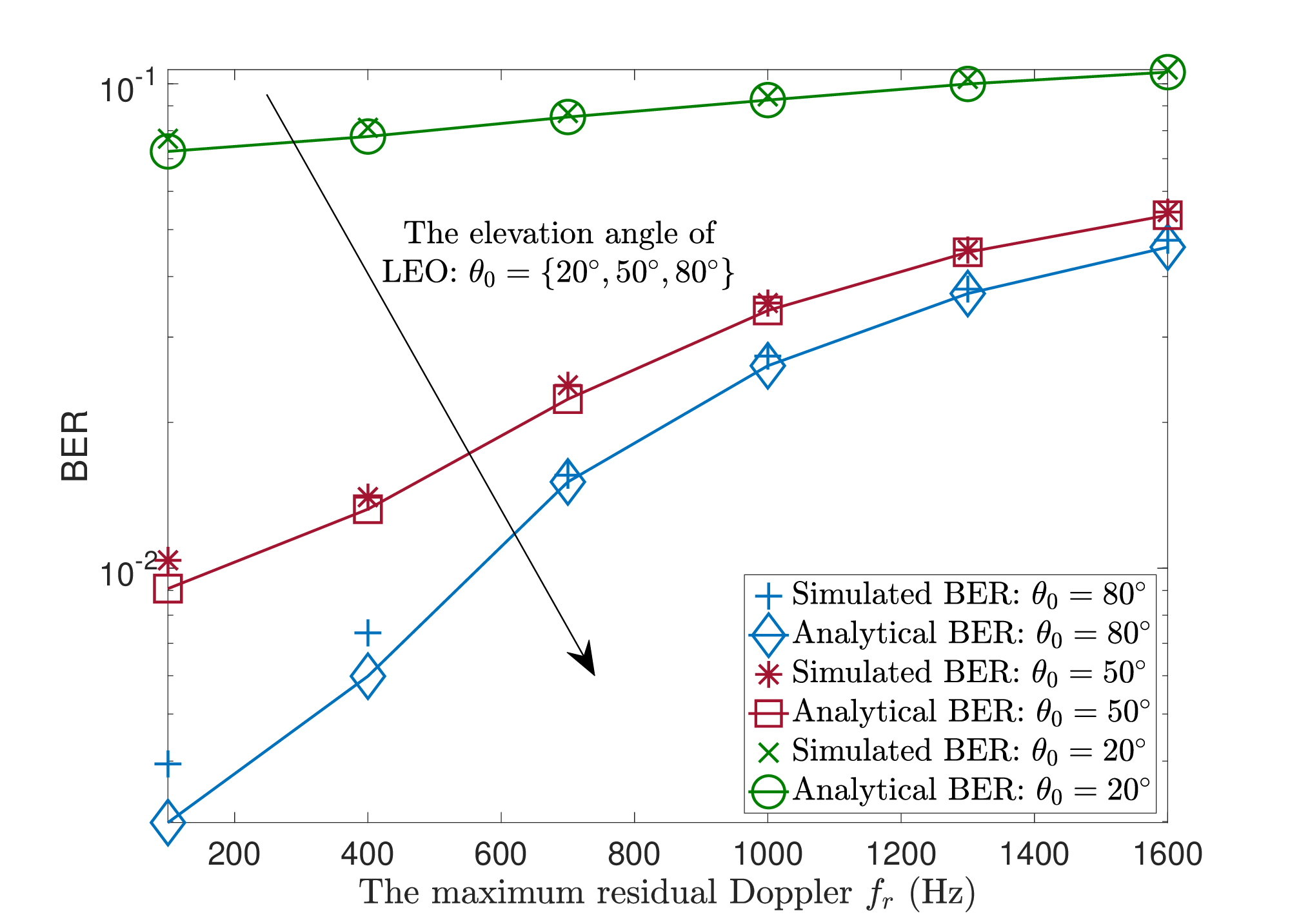}
    \caption{The BER performance versus the maximum residual Doppler $\bf{f_r}$ ($\bf{\left[100,1600\right]}$ Hz) with different elevation angles, denoted as $\bf{\theta_0 = \{20,50,80\}^{\circ}}$. }
        \label{fig:BER_fr_tau}
        \vspace{0.3cm}
\end{figure}

\begin{figure}[!htb]
\centering
    \includegraphics[width= 3.5 in]{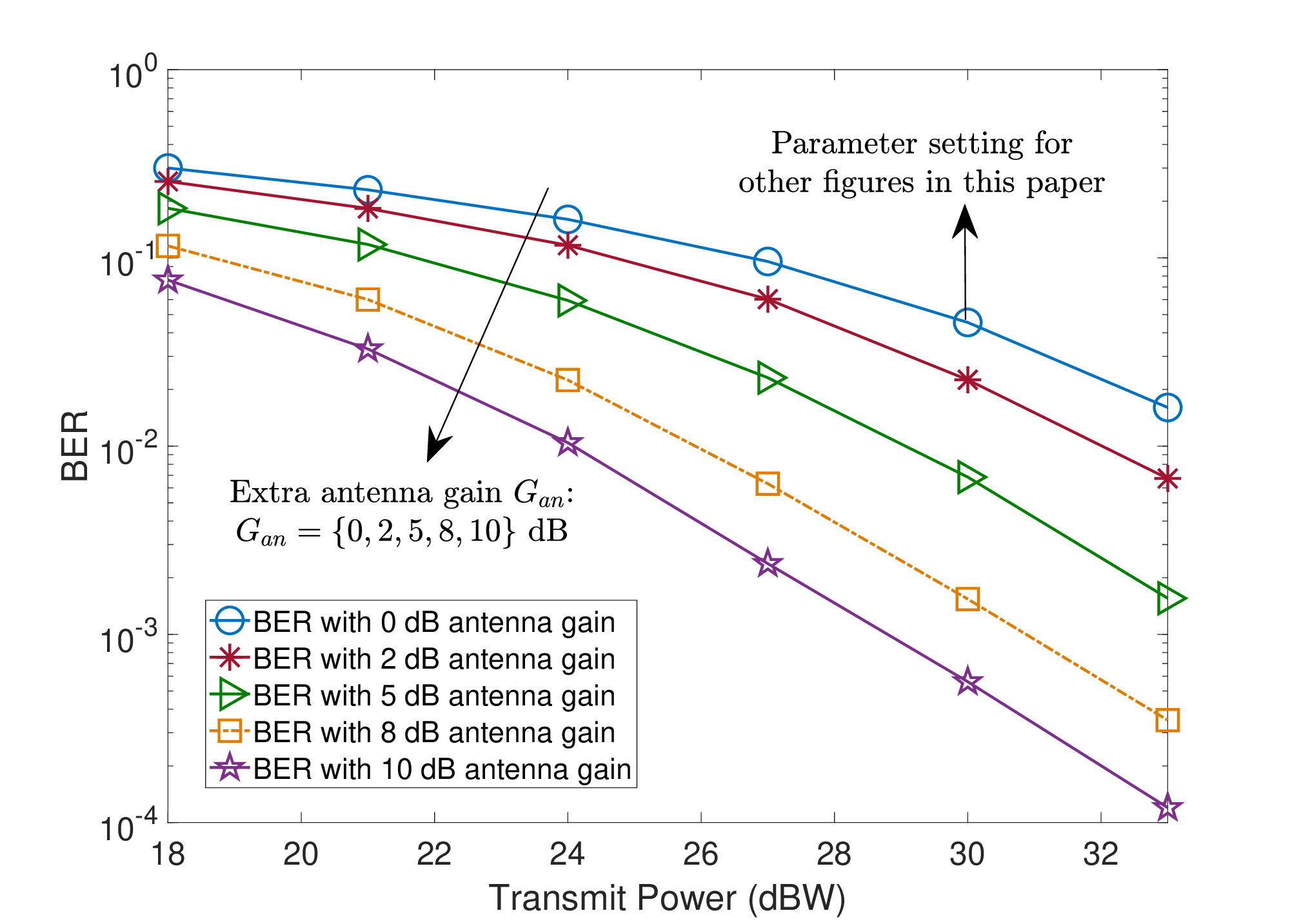}
    \caption{The BER performance versus the transmit power $\mathbf{P_w \in [18,33]}$ dBW, with additional antenna gain $\mathbf{G_{an} = \{0,2,5,8,10\}}$ dBi.}
        \label{fig:BER_TP}
\end{figure}

Fig. \ref{fig:BER_tau_Ps} portrays the BER performance versus the effective time delay between the pilot symbol and the data symbol, for the settings of $P_w =  36$ dBW, $K= 10$ dB, $m = 12$, $f_D = 100$ Hz, and $\tau_{e} \in \left[2 \times 10^{-4} , - 3\times 10^{-3}\right]$ s. We observe in Fig. \ref{fig:BER_tau_Ps} that the BER decreases when the time delay is reduced. Given the fact that the synchronization error may increase the time delay between the pilot and data symbols, this implies that imperfect time synchronization degrades the BER performance. The BER performance versus the maximum residual Doppler is visualized in Fig. \ref{fig:BER_fr_tau} for the settings of $P_w =  36$ dBW, $K= 10$ dB, $m = 12$, $\tau_{e} = 0.2$ ms, and $f_D \in \left[100:1600\right]$ Hz. Naturally, under a lower maximum residual Doppler, the BER performance is better than that under a higher Doppler. This is because a higher Doppler reduces the channel correlation between the pilot and data symbols, thereby introducing estimation and equalization errors. The figure also indicates that increasing the elevation angle significantly improves the BER performance owing to the reduced path loss. In satellite communications, the elevation angle typically increases from low to high and then decreases during a satellite passes above a terrestrial user, resulting in pronounced BER performance variations. This effect is particularly evident in the low-SNR regime, especially with synchronization errors. Hence, harnessing forward error correction or exploiting multiple antennas may be paramount in SAGINs.

Based on the previous simulation results (Fig. \ref{fig:BER_SNR_m} to \ref{fig:BER_fr_tau}), the BER observed tends to be above $10^{-2}$ even when the transmit power is high as 32 dBW. To further improve the BER performance, the effective antenna gain may be improved by multi-antenna beamforming techniques. Hence, Fig. \ref{fig:BER_TP} illustrates the BER performance versus the transmit power under different additional antenna gains (denoted as $G_{an}$). To elaborate, the blue line with circles represents the parameter settings of the previous figures without additional antenna gain, thus the BER performance is lower than that with antenna gain. This indicates that simply increasing the transmit power is not enough for SAGINs. Hence, by adding additional antenna gains by exploiting multiple antennas or optimizing beamforming desgins, it can be observed that increasing the additional antenna gain significantly improves the BER performance. This suggests that when the link budget of the satellite is inadequate, the BER can be readily improved by increasing the number of antennas along with optimal beamforming designs. By jointly considering Fig. \ref{fig:BER_fr_tau} and Fig. \ref{fig:BER_TP}, it may be observed that there exists a trade-off between hardware complexity (associated with multi-antenna systems) and computational complexity (introduced by channel coding techniques). This trade-off motivates further optimization efforts in SAGINs.

\section{Conclusions}
By considering the Doppler compensation errors and synchronization clock offsets, this paper has investigated the BER performance of time-varying channels encountered in SAGINs. Upon representing the residual Doppler by Jakes' Model, the distribution of the correlated Shadowed-Rician channel has been formulated as a bi-variate Gamma distribution with its correlation coefficient derived. Upon harnessing least-square channel estimation and equalization, we have derived the closed-form BER of 16-QAM in the face of imperfect Doppler compensation and synchronization, hence closing the related open problem in the SAGIN literature. Additionally, this paper provides physics-based theoretical insights on improving localization accuracy in ISaC systems. Since synchronization errors degrade system performance, future work will consider SAGINs employing forward error correction techniques and exploiting multiple antennas to improve performance.

\section*{APPENDIX~A: PROOF OF THEOREM \ref{T_E_orbit}}
\label{Appendix:A}
\renewcommand{\theequation}{A.\arabic{equation}}
\setcounter{equation}{0}

From Newton's law, we have the formula: $\ddot{\mathbf r} = -\frac{\mu}{r^3}\mathbf r $, with the angular momentum, formulated as $h = \left\lVert \mathbf r \times \dot{\mathbf r} \right\rVert$, and the specific mechanical energy, formulated as $\varepsilon = \frac{1}{2}\left\lVert\dot{\mathbf r}\right\rVert^2 - \frac{\mu}{r}$.

In the orbital plane, let $\dot{\theta} = h/r^2$. By exploiting $u(\theta)=1/r$, we can derive the resultant standard solution as:
\begin{align}
r(\theta)  = \frac{p}{1 + e\cos\theta}, \quad p =\frac{h^2}{\mu}.
\end{align}

Upon employing the relevant orbital geometry formulas, including $p = a(1-e^2)$ and $\varepsilon = -\frac{\mu}{2a}$, we obtain:
\begin{align}
a  = -\frac{\mu}{2\varepsilon}, \quad
e  = \sqrt{1 + \frac{2\varepsilon h^2}{\mu^2}}, \quad
b  = a\sqrt{1-e^2}.
\end{align}

Since we have the perigee/apogee expressions of $r_p=a(1-e)$ and $r_a=a(1+e)$, we can represent $a$ and $e$ by the perigee/apogee, formulated as:
\begin{align}
a &= \frac{r_p+r_a}{2}, \quad  e  = \frac{r_a-r_p}{r_a+r_p},
\end{align}
followed by obtaining $b = a\sqrt{1-e^2}$.

Let us define a pair of parameters, formulated as $x  = a\cos\varphi$ and $y  = b\sin\varphi$. Based on the geometric characteristics of the ellipse, we arrive at a differential equation, formulated as
$\frac{ds}{d\varphi}= \sqrt{a^2\sin^2\varphi + b^2\cos^2\varphi}= a\sqrt{1-e^2\cos^2\varphi}$. By integrating both sides of the above equation, we then arrive at the circumference of the ellipse, formulated as:
\begin{align}
C  = 4a \int_0^{\pi/2} \sqrt{1-e^2\cos^2\varphi}\, d\varphi  = 4a E(e).
\end{align}

\section*{APPENDIX~B: PROOF OF THEOREM \ref{relativity2}}
\label{Appendix:B}
\renewcommand{\theequation}{B.\arabic{equation}}
\setcounter{equation}{0}

Assuming a uniform Earth mass distribution and a locally inertial, non-rotating, freely falling coordinate system centered at the Earth's center of mass, we employ an approximate solution of Einstein's field equations in isotropic coordinates \cite{Ashby_Relativity_2003_Jan}. Starting from this metric, we factor out $(c\,dt)^2$ from the right-hand side, formulated as \eqref{B1}.

\begin{figure*}
\begin{align} \label{B1}
-ds^{2}
&=-\Bigg[1+\frac{2(V-\Phi_{0})}{c^{2}}-\left(1-\frac{2V}{c^{2}}\right)\frac{dr^{2}+r^{2}d\theta^{2}+r^{2}\sin^{2}\theta\,d\phi^{2}}{(c\,dt)^{2}}
\Bigg](c\,dt)^{2}.
\end{align}
\hrulefill
\end{figure*}

We then have the Earth-centered inertial (ECI) speed definition, formulated as:
\begin{align}
v^{2}=\frac{dr^{2}+r^{2}d\theta^{2}+r^{2}\sin^{2}\theta\,d\phi^{2}}{dt^{2}}.
\end{align}

Since we only retain the terms having the order of $c^{-2}$, the factor $\left(1-\frac{2V}{c^{2}}\right)$ multiplying the velocity term contributes only higher-order corrections $O(c^{-4})$, represented as:
\begin{align}
\frac{{d{s^2}}}{{{{(c\,dt)}^2}}} = 1 + \frac{{2(V - {\Phi _0})}}{{{c^2}}} - \frac{{{v^2}}}{{{c^2}}}+ O(c^{-4}).
\end{align}

Upon dropping the $O(c^{-4})$ terms, we have:
\begin{align}\label{c-4}
ds^{2}\simeq (c\,dt)^{2}\left[1+\frac{2(V-\Phi_{0})}{c^{2}}-\frac{v^{2}}{c^{2}}\right].
\end{align}

Taking the square root and expanding $\sqrt{1+\epsilon}\simeq 1+\epsilon/2$ for $\epsilon=O(c^{-2})$ yields:
\begin{align}
d\tau=\frac{ds}{c} &\simeq dt\left[1+\frac{V-\Phi_{0}}{c^{2}}-\frac{v^{2}}{2c^{2}}\right].
\end{align}

Finally, by exploiting $(1+\delta)^{-1}\simeq 1-\delta$ with $\delta=O(c^{-2})$, we have:
\begin{align}
dt
&\simeq d\tau\left[1-\frac{V-\Phi_{0}}{c^{2}}+\frac{v^{2}}{2c^{2}}\right],
\end{align}
and then integrate along the clock path to obtain the final expression for $\int_{\text{path}}dt$.

To compare the formula of special relativity, we could also directly derive the square root of \eqref{c-4}, yielding:
\begin{align}
dt
&\simeq d\tau \sqrt{ \left[1+\frac{2\left(V-\Phi_{0}\right)}{c^{2}}-\frac{v^{2}}{c^{2}}\right] },
\end{align}
which is exploited as the formula of \eqref{relativity_equation} in \textbf{Theorem \ref{relativity2}}.

\section*{APPENDIX~C: PROOF OF LEMMA \ref{L_Gaussian_C}}
\label{Appendix:C}
\renewcommand{\theequation}{C.\arabic{equation}}
\setcounter{equation}{0}

We first define the complex Gaussian fading process under Clarke-Jakes' model as:
\begin{equation} \label{Jakes_signal}
A(t) = \frac{1}{\sqrt{N}} \sum_{n=1}^{N} \left[X_n(t) + jY_n(t)\right],
\end{equation}
where $\{X_n(t)\}$ and $\{Y_n(t)\}$ are i.i.d.\ zero-mean real Gaussian processes with:
\begin{equation}
\mathbb{E}\!\left[X_n(t)X_m(t+\tau)\right] = \mathbb{E}\!\left[Y_n(t)Y_m(t+\tau)\right]= b_0 \rho_J(\tau)\,\delta_{nm},
\end{equation}
and $\delta_{nm}$ is the Kronecker delta, formulated as:
\begin{equation}
\delta_{nm} =
\begin{cases}
0, & \text{if } n \neq m,\\
1, & \text{if } n = m.
\end{cases}
\end{equation}

Since the NLoS portion (isotropic scattering scenario, namely the Rayleigh fading) has been defined as \eqref{complex_signal}, by aligning the parameters between \eqref{complex_signal} and \eqref{Jakes_signal}, we have:
\begin{align}
A_I(t) &=  \frac{1}{\sqrt{N}} \sum_{n=1}^{N} {X_n(t)},\\
A_Q(t) &= \frac{1}{\sqrt{N}} \sum_{n=1}^{N} {Y_n(t)},
\end{align}
hence $A_I(t)$ and $A_Q(t)$ are also i.i.d.\ Gaussian processes.

Then, the autocorrelation is formulated as:
\begin{align}
R_A(\tau) =& \mathbb{E}\!\left[A(t) A^\ast(t+\tau)\right] \notag \\
= &\mathbb{E}\!\left[A_I(t)A_I(t+\tau)\right]
+ \mathbb{E}\!\left[A_Q(t)A_Q(t+\tau)\right].
\end{align}

We denote the correlation matrix of $[X_1(t),\dots,X_N(t)]^\mathsf{T}$ as $\mathbf{R}_X(\tau)$ and that of $Y_1(t),\dots,Y_N(t)]^\mathsf{T}$ as $\mathbf{R}_Y(\tau)$. Hence, the correlation matrices and the autocorrelation expression are formulated as:
\begin{align}
\mathbf{R}_X(\tau)&=\mathbf{R}_Y(\tau)
= b_0 \rho_J(\tau)\mathbf{I}_N, \\
R_A(\tau)&= \frac{1}{N}\operatorname{tr}\!\left[\mathbf{R}_X(\tau)\right]+ \frac{1}{N}\operatorname{tr}\!\left[\mathbf{R}_X(\tau)\right]\notag\\
&= \frac{2}{N} \operatorname{tr}\!\left(b_0\rho_J(\tau)
\mathbf{I}_N\right) = 2 b_0 \rho_J(\tau),
\end{align}
where we have $\rho_J(\tau) = J_0(2\pi f_D\tau)$ and $R_{A_I}(\tau) = R_{A_Q}(\tau) = b_0 J_0(2\pi f_D\tau)$.

\section*{APPENDIX~D: PROOF OF LEMMA \ref{L_Gaussian_G}}
\label{Appendix:D}
\renewcommand{\theequation}{D.\arabic{equation}}
\setcounter{equation}{0}

This proof has four steps, derived in the following.

\medskip
\textbf{(1) Normalizing the Gamma representation}

To generate the Gamma process, we first define $2m$ normalized Gaussian processes, formulated as:
\begin{equation}
U_p(t) \triangleq X_p(t)\sqrt{\frac{2m}{\Omega}},\quad
p=1,\dots,2m,
\end{equation}
where we have $U_p(t)\sim\mathcal{N}(0,1)$ and
\begin{equation}
\mathbb{E}\!\left[U_p(t)U_p(t+\tau)\right] = \rho_J(\tau).
\end{equation}

Then, the pair of correlated Nakagami-$m$ processes are defined as:
\begin{equation}\label{D3}
S_1 \triangleq \sum_{p=1}^{2m} U_p^2(t),\quad
S_2 \triangleq \sum_{p=1}^{2m} U_p^2(t+\tau).
\end{equation}

Since the square of the Nakagami-$m$ process is a Gamma process, the correlated Gamma processes are formulated as:
\begin{equation}
Z^2(t) = \sum_{p=1}^{2m} X_p^2(t)
= \frac{\Omega}{2m} S_1,\quad
Z^2(t+\tau) = \frac{\Omega}{2m} S_2.
\end{equation}

Since $U_p(t)$ represents i.i.d. Gaussian processes, $S_1$ and $S_2$ follow the chi-square distribution (related to the Gamma distribution) with $2m$ degrees of freedom, formulated as:
\begin{equation}
S_i \sim \chi^2_{2m} \equiv \mathrm{Gamma}\!\left(m,2\right),\quad i=1,2.
\end{equation}

Hence, we define a pair of normalized Gamma variables, denoted as $G_1, G_2 \sim \mathrm{Gamma}(m,1)$ and formulated as:
\begin{equation}\label{D6}
G_1 \triangleq \frac{m}{\Omega}Z^2(t) = \frac{S_1}{2},\quad
G_2 \triangleq \frac{m}{\Omega}Z^2(t+\tau) = \frac{S_2}{2}.
\end{equation}

Thus, we have:
\begin{equation}
Z(t) = \sqrt{\frac{\Omega}{m}}\; G_1^{1/2},\quad
Z(t+\tau) = \sqrt{\frac{\Omega}{m}}\; G_2^{1/2}.
\end{equation}

The autocorrelation is expressed as:
\begin{align}\label{RZ}
R_Z(\tau)= \mathbb{E}[Z(t)Z(t+\tau)] = \frac{\Omega}{m}\,\mathbb{E}\!\left[ G_1^{1/2} G_2^{1/2} \right].
\end{align}

Compared to \eqref{RZ}, we obtain:
\begin{equation}\label{rhoZ_moment}
\rho_Z(\tau) = \frac{1}{m}\,\mathbb{E}\!\left[ G_1^{1/2} G_2^{1/2} \right].
\end{equation}

In the next step, the moment $\mathbb{E}[G_1^{1/2} G_2^{1/2}]$ for the correlated Gamma pair $(G_1,G_2)$ generated by the underlying correlated Gaussian processes has to be derived.

\medskip
\textbf{(2) Joint PDF of correlated Gamma variables}

To generate the joint PDF of the correlated Gamma variables, we define a pair of vectors, whose elements are zero-mean Gaussian variables, formulated as:
\begin{align}
\mathbf{U}(t) & = [U_1(t),\dots,U_{2m}(t)]^\mathsf{T},\\
\mathbf{U}(t+\tau) &= [U_1(t+\tau),\dots,U_{2m}(t+\tau)]^\mathsf{T}.
\end{align}

Hence, the covariance matrices of the above two vectors are:
\begin{align}
&\mathbb{E}\!\big[\mathbf{U}(t)\otimes \mathbf{U}^\mathsf{T}(t)\big]
= \mathbb{E}\!\big[\mathbf{U}(t+\tau)\otimes \mathbf{U}^\mathsf{T}(t+\tau)\big] = \mathbf{I}_{2m},\quad \\
&\mathbb{E}\!\big[\mathbf{U}(t)\otimes \mathbf{U}^\mathsf{T}(t+\tau)\big]
= \rho_J(\tau)\,\mathbf{I}_{2m}.
\end{align}

Equivalently, each pair $(U_p(t),U_p(t+\tau))$ obeys i.i.d.\ bi-variate Gaussian processes with their correlation coefficients, equaling $\rho_J(\tau)$. By substituting \eqref{D3} into \eqref{D6}, we have:
\begin{equation}
G_1 = \frac{1}{2}\sum_{p=1}^{2m} U_p^2(t),\quad
G_2 = \frac{1}{2}\sum_{p=1}^{2m} U_p^2(t+\tau).
\end{equation}

Thus, the joint PDF of $(G_1,G_2)$ may be given by the bi-variate Gamma density function, formulated as:
\begin{align}\label{biv_gamma_pdf}
& {f_{{G_1},{G_2}}}(x,y) = \frac{{{m^{m + 1}}{(xy)^{m - 1}} }}{{\Gamma (m){{(1 - \rho _J^2(\tau ))}^m}}}\exp \left( { - \frac{{m(x + y)}}{{1 - \rho _J^2(\tau )}}} \right)\notag\\
 &\hspace*{0.2cm} \times {I_{m - 1}}\left( {\frac{{2m\,{\rho _J}(\tau )\sqrt {xy} }}{{1 - \rho _J^2(\tau )}}} \right),\quad x > 0,\:y > 0,
\end{align}
where $I_{\nu}(\cdot)$ is the modified Bessel function of the first kind and order $\nu$.

\begin{figure*}[hpt!]
\begin{align} \label{Average_aa}
 {\mathbb{E}}\left[ {{\alpha ^2}{{\hat \alpha }^2}} \right] %=& {\mathbb{E}}\left[ {{{\left| {{h_{ST}}\left( {t + \tau_{e} } \right)} \right|}^2}{{\left| {{h_{ST}}\left( t \right) } \right|}^2}} \right]  \notag \\
 = & \mathbb{E}\left[ {\left( {A_I^2\left( {t + \tau_{e} } \right) + A_Q^2\left( {t + \tau_{e} } \right) + {Z^2}\left( {t + \tau_{e} } \right) + 2{A_I}\left( {t + \tau_{e} } \right)Z\left( {t + \tau_{e} } \right)\sin \xi  + 2{A_Q}\left( {t + \tau_{e} } \right)Z\left( {t + \tau_{e} } \right)\cos \xi } \right)} \right.\notag \\
&  \times \left. {\left( {A_I^2\left( t \right) + A_Q^2\left( t \right) + {Z^2}\left( t \right) + 2{A_I}\left( t \right)Z\left( t \right)\sin \xi  + 2{A_Q}\left( t \right)Z\left( t \right)\cos \xi } \right)} \right]  \notag \\
= & 4b_0^2\left( {\rho _J^2\left( \tau_{e}  \right) + 1} \right) + 4{b_0}\Omega  + {\Omega ^2}\left( {1 + \frac{{\rho _J^2\left( \tau_{e}  \right)}}{m}} \right) + 4{b_0}\Omega {\rho _J}\left( \tau_{e}  \right){\rho _Z}\left( \tau_{e}  \right) . \tag{E.4}
\end{align}
\hrulefill
\end{figure*}

\begin{figure*}[hpt!]
\begin{align}\label{A3}
&{\mathbb{E}}\left[ {{\alpha ^4}} \right]  =  {\mathbb{E}}\left[ {{{\left| {{h_{ST}}\left( {t + \tau_{e} } \right)} \right|}^2}{{\left| {{h_{ST}}\left( {t + \tau_{e} } \right)} \right|}^2}} \right] ={\mathbb{E}}\left[ {{{\left( {A_I^2 + A_Q^2 + {Z^2} + 2{A_I}Z\sin \xi  + 2{A_Q}Z\cos \xi } \right)}^2}} \right] \notag\\
&  = {\mathbb{E}}\left[ {A_I^2\left( {A_I^2 + A_Q^2 + {Z^2} + 2{A_I}Z\sin \xi  + 2{A_Q}Z\cos \xi } \right)} \right] + {\mathbb{E}}\left[ {A_Q^2\left( {A_I^2 + A_Q^2 + {Z^2} + 2{A_I}Z\sin \xi  + 2{A_Q}Z\cos \xi } \right)} \right] \notag\\
 & + {\mathbb{E}}\left[ {{Z^2}\left( {A_I^2 + A_Q^2 + {Z^2} + 2{A_I}Z\sin \xi  + 2{A_Q}Z\cos \xi } \right)} \right]   + {\mathbb{E}}\left[ {2{A_I}Z\sin \xi \left( {A_I^2 + A_Q^2 + {Z^2} + 2{A_I}Z\sin \xi  + 2{A_Q}Z\cos \xi } \right)} \right] \notag\\
 & + {\mathbb{E}}\left[ {2{A_Q}Z\cos \xi \left( {A_I^2 + A_Q^2 + {Z^2} + 2{A_I}Z\sin \xi  + 2{A_Q}Z\cos \xi } \right)} \right] \notag\\
 & = \underbrace {2{\mathbb{E}}\left[ {A_I^4} \right]}_{ = 6b_0^2} + 2b_0^2 + 8{b_0}\Omega  + {\mathbb{E}}\left[ {{Z^4}} \right] + \underbrace {4\left( {\sin \xi  + \cos \xi } \right){\mathbb{E}}\left[ {A_I^3Z} \right]}_{ = 0}  = 8b_0^2 + 8{b_0}\Omega  + {\Omega ^2}\left( {1 + \frac{1}{m}} \right).  \tag{E.7}
\end{align}
\hrulefill
\end{figure*}

\medskip
\textbf{(3) Mixed moment $\mathbb{E}[G_1^{1/2} G_2^{1/2}]$}

Based on \eqref{biv_gamma_pdf}, we derive the following integrals:
\begin{align} \label{moment_integral}
&\mathbb{E}\!\left[G_1^{1/2}G_2^{1/2}\right]
= \int_0^\infty\!\!\int_0^\infty x^{1/2}y^{1/2} f_{G_1,G_2}(x,y)\,dx\,dy \notag\\
& = \frac{m^{m+1}}{\Gamma(m)\big(1-\rho_J^2(\tau)\big)^m} \int_0^\infty\!\!\int_0^\infty x^{m-\tfrac{1}{2}}y^{m-\tfrac{1}{2}} \notag\\
& \times \exp\!\left(-\frac{m(x+y)}{1-\rho_J^2(\tau)}\right) I_{m-1}\!\left(\frac{2m\,\rho_J(\tau)\sqrt{xy}}{1-\rho_J^2(\tau)}\right)\,dx\,dy.
\end{align}

By substituting  the series expansion of the modified Bessel function formulated as $I_{m-1}(z) = \sum_{k=0}^{\infty} \frac{1}{k!\,\Gamma(k+m)} \left(\frac{z}{2}\right)^{2k+m-1}$ into \eqref{moment_integral} yields:
\begin{align}\label{moment_series}
&\mathbb{E}\!\left[G_1^{1/2}G_2^{1/2}\right]
= \frac{m^{m+1}}{\Gamma(m)\big(1-\rho_J^2(\tau)\big)^m}
\sum_{k=0}^{\infty}\frac{\left(\frac{m\,\rho_J(\tau)}{1-\rho_J^2(\tau)}\right)^{2k+m-1}}{k!\,\Gamma(k+m)}
 \notag\\
&\hspace*{0.7cm} \times \int_0^\infty x^{m+k-\tfrac{1}{2}}
\exp\!\left(-\frac{m}{1-\rho_J^2(\tau)}x\right)dx \notag\\
&\hspace*{0.7cm}\times \int_0^\infty y^{m+k-\tfrac{1}{2}}
\exp\!\left(-\frac{m}{1-\rho_J^2(\tau)}y\right)dy.
\end{align}

In \eqref{moment_series} which each integral is a Gamma integral, obeying the form of:
\begin{align}
\int_0^\infty x^{m+k-\tfrac{1}{2}}
\exp\!\left(-\frac{m}{1-\rho_J^2(\tau)}x\right)dx \notag\\
\hspace*{0.5cm}= \Gamma\!\left(m+k+\tfrac{1}{2}\right)
\left(\frac{1-\rho_J^2(\tau)}{m}\right)^{m+k+\tfrac{1}{2}}.
\end{align}

Thus, equation \eqref{moment_series} becomes:
\begin{align}
&\mathbb{E}\!\left[G_1^{1/2}G_2^{1/2}\right]
= \frac{m^{m+1}}{\Gamma(m)\big(1-\rho_J^2(\tau)\big)^m}  \notag\\
&\quad \times \sum_{k=0}^{\infty}\frac{1}{k!\,\Gamma(k+m)}
\left(\frac{m\,\rho_J(\tau)}{1-\rho_J^2(\tau)}\right)^{2k+m-1} \notag\\
&\quad \times \Gamma^2\!\left(m+k+\tfrac{1}{2}\right)
\left(\frac{1-\rho_J^2(\tau)}{m}\right)^{2m+2k+1} \notag\\
&= \frac{1}{\Gamma(m)}\left(\frac{1-\rho_J^2(\tau)}{m}\right) \sum_{k=0}^{\infty}
\frac{\Gamma^2\!\left(m+k+\tfrac{1}{2}\right)}{\Gamma(k+m)\,k!}
\left(\rho_J^2(\tau)\right)^k. \label{moment_series2}
\end{align}

We now employ the Pochhammer symbol, defined as $(a)_k = \Gamma(a+k)/\Gamma(a)$, and arrive at $\Gamma\!\left(m+k+\tfrac{1}{2}\right) = \Gamma\!\left(m+\tfrac{1}{2}\right)\left(m+\tfrac{1}{2}\right)_k$ and $\Gamma(k+m) = \Gamma(m)\,(m)_k$. Then, equation \eqref{moment_series2} is further formulated as:
\begin{align}
\mathbb{E}\!\left[G_1^{1/2}G_2^{1/2}\right]
&= \frac{1}{\Gamma(m)}\left(\frac{1-\rho_J^2(\tau)}{m}\right)
\Gamma^2\!\left(m+\tfrac{1}{2}\right) \notag\\
&\quad \times \sum_{k=0}^{\infty}
\frac{\big(m+\tfrac{1}{2}\big)_k^2}{(m)_k\,k!}
\left(\rho_J^2(\tau)\right)^k. \label{moment_pochhammer}
\end{align}

By exploiting the definition of the hypergeometric series, formulated as $ {}_2F_1(a,b;c;z) = \sum_{k=0}^{\infty} \frac{(a)_k (b)_k}{(c)_k\,k!}\, z^k $, setting $a=b=-\tfrac{1}{2}$ and $c=m$, as well as using the identity of $\left(m+\tfrac{1}{2}\right)_k =\frac{\Gamma\!\left(m+\tfrac{1}{2}+k\right)} {\Gamma\!\left(m+\tfrac{1}{2}\right)}$, the final momentum may be expressed in the compact form of:
\begin{equation}\label{moment_final}
\mathbb{E}\!\left[G_1^{1/2}G_2^{1/2}\right] {=} \left(\frac{\Gamma\!\left(m{+}\tfrac{1}{2}\right)}{\Gamma(m)}\right)^2{}_2F_1\!\left({-}\frac{1}{2},{-}\frac{1}{2};\,m;\,\rho_J^2(\tau)\right).
\end{equation}

\medskip
\textbf{(4) The final expression of $\rho_Z(\tau)$}

Upon substituting \eqref{moment_final} into \eqref{rhoZ_moment}, we obtain:
\begin{equation}\label{moment_correlationZ}
\rho_Z(\tau)= \frac{1}{m}\left(\frac{\Gamma\!\left(m+\tfrac{1}{2}\right)} {\Gamma(m)}\right)^2{}_2F_1\!\left(-\frac{1}{2},-\frac{1}{2};\,m;\,\rho_J^2(\tau)\right).
\end{equation}

Then, by comparing \eqref{moment_correlationZ} and \eqref{RZ}, the autocorrelation may be expressed as
$R_Z(\tau) = \Omega \,\rho_Z(\tau).$

\section*{APPENDIX~E: PROOF OF THEOREM \ref{T_coefficient}}
\label{Appendix:E}
\renewcommand{\theequation}{E.\arabic{equation}}
\setcounter{equation}{0}

Upon recalling that the correlation coefficient is expressed as
${\rho _{SR}}\left( \tau_{e}  \right) = \frac{{{\mathop{\rm cov}} \left\{ {{\alpha ^2},{{\hat \alpha }^2}} \right\}}}{{\sqrt {{\mathop{\rm var}} \left( {{\alpha ^2}} \right){\mathop{\rm var}} \left( {{{\hat \alpha }^2}} \right)} }}$, we will now derive three formulas one by one, including ${\mathop{\rm cov}} \left\{ {{\alpha ^2},{{\hat \alpha }^2}} \right\} $, ${\mathop{\rm var}} \left( {{\alpha ^2}} \right)$, and ${\mathop{\rm var}} \left( {{{\hat \alpha }^2}} \right)$.

\medskip
\textbf{(1) Derivation of ${\mathop{\rm cov}} \left\{ {{\alpha ^2},{{\hat \alpha }^2}} \right\} $}

Given ${\mathop{\rm cov}} \left\{ {{\alpha ^2},{{\hat \alpha }^2}} \right\} = {\mathbb{E}}\left[ {{\alpha ^2}{{\hat \alpha }^2}} \right] - {\mathbb{E}}\left[ {{\alpha ^2}} \right]{\mathbb{E}}\left[ {{{\hat \alpha }^2}} \right]$, we separately calculate ${\mathbb{E}}\left[ {{\alpha ^2}} \right]$ and ${\mathbb{E}}\left[ {{{\hat \alpha }^2}} \right]$ as:
\begin{align}\label{A1}
& {\mathbb{E}}\left[ {{\alpha ^2}} \right] = {\mathbb{E}}\left[ {{{\left| {{h_{ST}}\left( {t + \tau_{e} } \right)} \right|}^2}} \right]  = 2{b_0} + \Omega,  \\
\label{A2}
& {\mathbb{E}}\left[ {{{\hat \alpha }^2}} \right] = {\mathbb{E}}\left[ {{{\left| {{{\hat h}_{ST}}\left( t \right)} \right|}^2}} \right] = 2{b_0} + \Omega.
\end{align}

The expectation ${\mathbb{E}}\left[ {{\alpha ^2}{{\hat \alpha }^2}} \right]$ is detailed in \eqref{Average_aa} based on the following formulas, expressed as:
\begin{align}
& \mathbb{E} \left[ {A_I^2\left( {t + {\tau _e}} \right)A_I^2\left( t \right)} \right] = \mathbb{E}\left[ {A_Q^2\left( {t + {\tau _e}} \right)A_Q^2\left( t \right)} \right]\notag\\
&\hspace*{3.2cm}  = b_0^2\left[ {1 + 2\rho _J^2\left( {{\tau _e}} \right)} \right], \\
& \mathbb{E}\left[ {A_I^2\left( {t + {\tau _e}} \right)A_Q^2\left( t \right)} \right] =\mathbb{E} \left[ {A_Q^2\left( {t + {\tau _e}} \right)A_I^2\left( t \right)} \right] = b_0^2, \\
& \mathbb{E}\left[ {A_I^2\left( {t + {\tau _e}} \right){Z^2}\left( t \right)} \right] = \mathbb{E}\left[ {A_Q^2\left( {t + {\tau _e}} \right){Z^2}\left( t \right)} \right] = {b_0}\Omega , \\
 &\mathbb{E}\left[ {{Z^2}\left( {t + {\tau _e}} \right)A_I^2\left( t \right)} \right] =\mathbb{E} \left[ {{Z^2}\left( {t + \tau } \right)A_Q^2\left( t \right)} \right] = {b_0}\Omega , \\
& \mathbb{E} \left[ {{Z^2}\left( {t + {\tau _e}} \right){Z^2}\left( t \right)} \right] = \rho _J^2\left( {{\tau _e}} \right)\frac{{{\Omega ^2}}}{m} + {\Omega ^2}, \\
& \mathbb{E} \left[ {{A_I}\left( {t + {\tau _e}} \right)Z\left( {t + {\tau _e}} \right){A_I}\left( t \right)Z\left( t \right)} \right] = {b_0}\Omega {\rho _J}\left( {{\tau _e}} \right){\rho _Z}\left( {{\tau _e}} \right), \\
& \mathbb{E} \left[ {{A_Q}\left( {t + {\tau _e}} \right)Z\left( {t + {\tau _e}} \right){A_Q}\left( t \right)Z\left( t \right)} \right] {=} {b_0}\Omega {\rho _J}\left( {{\tau _e}} \right){\rho _Z}\left( {{\tau _e}} \right) .
 \end{align}

By substituting \eqref{A1}, \eqref{A2}, and \eqref{Average_aa} into the definition of ${\mathop{\rm cov}} \left\{ {{\alpha ^2},{{\hat \alpha }^2}} \right\}$, the crosscorrelation is formulated as:
\begin{align}
 &{\mathop{\rm cov}} \left\{ {{\alpha ^2},{{\hat \alpha }^2}} \right\} = {\rm{E}}\left[ {{\alpha ^2}{{\hat \alpha }^2}} \right] - {\rm{E}}\left[ {{\alpha ^2}} \right]{\rm{E}}\left[ {{{\hat \alpha }^2}} \right] \notag \\
 & = \rho _J^2\left( \tau_{e}  \right)4b_0^2 + \frac{{{\Omega ^2}}}{m}\rho _J^2\left( \tau_{e}  \right) + 4{b_0}\Omega {\rho _J}\left( \tau_{e}  \right){\rho _Z}\left( \tau_{e}  \right).
  \end{align}

\medskip
\textbf{(2) Derivation of ${\mathop{\rm var}} \left( {{\alpha ^2}} \right)$ and ${\mathop{\rm var}} \left( {{{\hat \alpha }^2}} \right)$}

These variances are defined as ${\mathop{\rm var}} \left( {{\alpha ^2}} \right) = {\mathbb{E}}\left[ {{\alpha ^4}} \right] - {\left( {{\mathbb{E}}\left[ {{\alpha ^2}} \right]} \right)^2}$ and ${\rm{ var}} \left( {{{\hat \alpha }^2}} \right) = {\mathbb{E}}\left[ {{{\hat \alpha }^4}} \right] - {\left( {{\mathbb{E}}\left[ {{{\hat \alpha }^2}} \right]} \right)^2}$. Hence, we have to derive ${\mathbb{E}}\left[ {{\alpha ^4}} \right]$ and ${\mathbb{E}}\left[ {{{\hat \alpha }^4}} \right]$, since we have derived ${{\mathbb{E}}\left[ {{\alpha ^2}} \right]}$ and ${{\mathbb{E}}\left[ {{{\hat \alpha }^2}} \right]}$ as \eqref{A1} and \eqref{A2}.

\setcounter{equation}{10}

The expectation of ${\mathbb{E}}\left[ {{\alpha ^4}} \right]$ is calculated as \eqref{A3} and ${\mathbb{E}}\left[ {{\hat \alpha ^4}} \right]$ is formulated as:
\begin{align}
& {\mathbb{E}}\left[ {{{\hat \alpha }^4}} \right] ={\mathbb{E}}\left[ {{{\left| {{h_{ST}}\left( {t  } \right)} \right|}^2}{{\left| {{h_{ST}}\left( {t  } \right)} \right|}^2}} \right] \notag\\
 &={\mathbb{E}}\left[ {{{\left| {{h_{ST}}\left( {t + \tau_{e} } \right)} \right|}^2}{{\left| {{h_{ST}}\left( {t + \tau_{e} } \right)} \right|}^2}} \right]
={\mathbb{E}}\left[ {{\alpha ^4}} \right].
\end{align}

Hence, the variances of ${\alpha ^2}$ and ${\hat \alpha }^2$ are derived as:
\begin{align}
\label{var1}
{\mathop{\rm var}} \left( {{\alpha ^2}} \right) &= 4b_0^2 + 6{b_0}\Omega  + \frac{{{\Omega ^2}}}{m},\\
\label{var2}
{\mathop{\rm var}} \left( {{{\hat \alpha }^2}} \right) &= {\mathop{\rm var}} \left( {{\alpha ^2}} \right)  = 4b_0^2 + 6{b_0}\Omega  + \frac{{{\Omega ^2}}}{m}.
\end{align}

Finally, by substituting \eqref{Average_aa}, \eqref{var1} and \eqref{var2} into \eqref{T_coefficient}, this proof is concluded.

\section*{APPENDIX~F: PROOF OF THEOREM \ref{T_BER_MQAM}}
\label{Appendix:F}
\renewcommand{\theequation}{F.\arabic{equation}}
\setcounter{equation}{0}

Upon substituting the parameters from $\{x,y\}$ to $\{r,\theta\}$ into \eqref{nakagami}, the bi-variate PDF of $\alpha$ and $\hat \alpha$ is derived as:
\begin{align}
 &{f_{\alpha ,\hat \alpha }}\left( {r,\theta ;{\gamma _1},{\gamma _2},\eta } \right) = \frac{{4{{\cos }^{2{\gamma _1}}}\left( \theta  \right){{\tan }^{{\gamma _2}}}\left( \theta  \right)}}{{\Gamma \left( {{\gamma _1}} \right)\Gamma \left( {{\gamma _2} - {\gamma _1}} \right)}}\notag \\
&\hspace*{0.5cm}\times  {r^{2{\gamma _1} + 1}}\exp \left( { - {r^2}} \right)\sum\limits_{k = 0}^\infty  {\frac{{{{\left( {1 - \eta } \right)}^{{\gamma _2}}}}}{{{\eta ^{\frac{{{\gamma _2} - k - 1}}{2}}}}}} \frac{{\Gamma \left( {{\gamma _2} - {\gamma _1} + k} \right)}}{{k!}} \notag\\
&\hspace*{0.5cm} \times {\tan ^k}\left( \theta  \right){I_{{\gamma _2} + k - 1}}\left( {\sqrt \eta  \sin \left( {2\theta } \right){r^2}} \right) .
\end{align}

With the aid of the Chebyshev-Gaussian Quadrature, the Gaussian $Q$ function is expressed as:
\begin{align}
 &Q\left( {zx} \right) = \frac{1}{\pi }\int_0^{\frac{\pi }{2}} {\exp \left( { - \frac{{{z^2}\left( {1 - \eta } \right){{\cos }^2}\left( \theta  \right)}}{{2{\beta _1}{{\sin }^2}t}}{r^2}} \right)dt} \notag  \\
  = &\frac{1}{4}\sum\limits_{i = 1}^Q {{w_i}} \sqrt {1 - t_i^2} \exp \left( { - \frac{{{z^2}\left( {1 - \eta } \right){{\cos }^2}\left( \theta  \right)}}{{2{\beta _1}{{\sin }^2}{\kappa _i}}}{r^2}} \right).
\end{align}

Then, the integral $\Upsilon \left( {z,0} \right)$ is expressed as:
\begin{align}\label{Upsilon1}
& \Upsilon \left( {z,0} \right) = \int_0^{\frac{\pi }{2}} {\sum\limits_{i = 1}^Q {{w_i}} \sqrt {1 - t_i^2} } \sum\limits_{k = 0}^\infty  {{{\cos }^{2{\gamma _1}}}\left( \theta  \right)} \notag \\
&\hspace*{1.3cm} \times {\tan ^{{\gamma _2} + k}}\left( \theta  \right)\frac{{{{\left( {1 - \eta } \right)}^{{\gamma _2}}}\Gamma \left( {{\gamma _2} - {\gamma _1} + k} \right)}}{{{\eta ^{\frac{{{\gamma _2} - k - 1}}{2}}}\Gamma \left( {{\gamma _1}} \right)\Gamma \left( {{\gamma _2} - {\gamma _1}} \right)k!}}d\theta \notag \\
&\times \underbrace {\int_0^\infty  {{r^{2{\gamma _1} + 1}}} \exp \left( { - f\left( {{\kappa _i},\theta ,z} \right){r^2}} \right){I_{{\gamma _2} + k - 1}}\left( {g\left( \theta  \right){r^2}} \right)dr}_{{\Xi _1}\left( {{\kappa _i},\theta ,z} \right)}.
\end{align}

Given the integral equation, formulated as:
\begin{align} \label{integral}
&  \int_0^\infty  {{x^\beta }} {e^{ - a{x^2}}}{I_\mu }(\gamma {x^2})\,dx =\frac{{{\gamma ^\mu }\Gamma \left[ {\frac{1}{2}(\beta  + 2\mu  + 1)} \right]}}{{{2^{\mu  + 1}}{a^{\frac{1}{2}(\beta  + 2\mu  + 1)}}\Gamma (\mu  + 1)}} \notag \\
& \times  {}_2{F_1}\left( {\frac{1}{4}(\beta  {+} 2\mu  {+} 1),\frac{1}{4}(\beta  {+} 2\mu  {+} 3),\mu  {+} 1;\frac{{{\gamma ^2}}}{{{a^2}}}} \right),
\end{align}
we have ${\Xi _1}\left( {{\kappa _i},\theta ,z} \right)$, derived as:
\begin{align} \label{Xi1}
& {\Xi _1}\left( {{\kappa _i},\theta ,z} \right) = \frac{{\Gamma \left( {{\gamma _1} + {\gamma _2} + k} \right)}}{{{2^{{\gamma _2} + k}}\Gamma \left( {{\gamma _2} + k} \right)}}\frac{{{g^{{\gamma _2} + k - 1}}\left( \theta  \right)}}{{{f^{{\gamma _1} + {\gamma _2} + k}}\left( {{\kappa _i},\theta ,z} \right)}}\notag  \\
 &\times {}_2{F_1}\left( {\frac{{{\gamma _1} + {\gamma _2} {+} k}}{2},\frac{{{\gamma _1} {+} {\gamma _2} {+} k {+} 1}}{2},{\gamma _2} + k;\frac{{{g^2}\left( \theta  \right)}}{{{f^2}\left( {{\kappa _i},\theta ,z} \right)}}} \right).
\end{align}

Then, by substituting \eqref{Xi1} into \eqref{Upsilon1} and exploiting the Chebyshev-Gauss quadrature, the integral of $\Upsilon \left( {z,0} \right) $ is derived as \eqref{U1}.

Following similar procedure to that of deriving $\Upsilon \left( {z,0} \right) $, we replace the Gaussian $Q$ function by its Chebyshev-Gauss series before deriving $\Upsilon \left( {z,\zeta} \right)$. Then, the function $\Upsilon \left( {z,\zeta} \right)$ is formulated as:
\begin{align} \label{Upsilon2}
&  {\Upsilon _2}\left( {z,\zeta } \right) = \int_0^{\frac{\pi }{2}} {\sum\limits_{i = 1}^Q {{w_i}} \sqrt {1 - t_i^2} } {\cos ^{2{\gamma _1}}}\left( \theta  \right){\tan ^{{\gamma _2}}}\left( \theta  \right) \notag \\
& \hspace*{1.5cm} \times \sum\limits_{k = 0}^\infty  {\frac{{{{\left( {1 - \eta } \right)}^{{\gamma _2}}}\Gamma \left( {{\gamma _2} - {\gamma _1} + k} \right){{\tan }^k}\theta }}{{{\eta ^{\frac{{{\gamma _2} - k - 1}}{2}}}\Gamma \left( {{\gamma _1}} \right)\Gamma \left( {{\gamma _2} - {\gamma _1}} \right)k!}}d\theta } \notag \\
& \small{{\times} {\underbrace {\int_0^\infty  {{r^{2{\gamma _1} {+} 1}}} \exp \left( { {-} {f_2}\left( {{\kappa _i},\theta ,z,\zeta } \right){r^2}} \right){I_{{\gamma _2} {+} k {-} 1}}\left( {g\left( \theta  \right){r^2}} \right)dr}_{{\Xi _2}\left( {{\kappa _i},\theta ,z,\zeta } \right)}.}}
\end{align}

Subsequently, based on \eqref{integral}, the integral of $\Xi _2$ is formulated as:
\begin{align} \label{Xi2}
& {\Xi _2}\left( {{\kappa _i},\theta ,z,\zeta } \right) = \frac{{\Gamma \left( {{\gamma _1} + {\gamma _2} + k} \right)}}{{{2^{{\gamma _2} + k}}\Gamma \left( {{\gamma _2} + k} \right)}}\frac{{{g^{{\gamma _2} + k - 1}}\left( \theta  \right)}}{{f_2^{{\gamma _1} + {\gamma _2} + k}\left( {{\kappa _i},\theta ,z,\zeta } \right)}}  \notag \\
&  \times  {}_2{F_1}\left( {\frac{{{\gamma _1} {+} {\gamma _2} {+} k}}{2},\frac{{{\gamma _1} {+} {\gamma _2} {+} k {+} 1}}{2},{\gamma _2} {+} k;\frac{{{g^2}\left( \theta  \right)}}{{f_2^2\left( {{\kappa _i},\theta ,z,\zeta } \right)}}} \right).
\end{align}

Finally, by substituting \eqref{Xi2} into \eqref{Upsilon2}, the derivation of $\Upsilon \left( {z,\zeta} \right) $ is completed and the proof is concluded.

{\setstretch{1}
\bibliographystyle{IEEEtran}
\bibliography{IEEEabrv,mybib}

% Generated by IEEEtran.bst, version: 1.13 (2008/09/30)
\begin{thebibliography}{10}
\providecommand{\url}[1]{#1}
\csname url@samestyle\endcsname
\providecommand{\newblock}{\relax}
\providecommand{\bibinfo}[2]{#2}
\providecommand{\BIBentrySTDinterwordspacing}{\spaceskip=0pt\relax}
\providecommand{\BIBentryALTinterwordstretchfactor}{4}
\providecommand{\BIBentryALTinterwordspacing}{\spaceskip=\fontdimen2\font plus
\BIBentryALTinterwordstretchfactor\fontdimen3\font minus
  \fontdimen4\font\relax}
\providecommand{\BIBforeignlanguage}[2]{{%
\expandafter\ifx\csname l@#1\endcsname\relax
\typeout{** WARNING: IEEEtran.bst: No hyphenation pattern has been}%
\typeout{** loaded for the language `#1'. Using the pattern for}%
\typeout{** the default language instead.}%
\else
\language=\csname l@#1\endcsname
\fi
#2}}
\providecommand{\BIBdecl}{\relax}
\BIBdecl

\bibitem{rohde_schwarz_5GAdv_NTN}
\BIBentryALTinterwordspacing
{Rohde \& Schwarz}. (2025) {5G-Adv. NTN technology for direct satellite access
  - from smart phone to automotive}. Rohde \& Schwarz Knowledge+. Video,
  speaker: I-Kang Fu, Senior Director Technology, MediaTek. [Online].
  Available:
  \url{https://www.rohde-schwarz.com/us/knowledge-center/videos/5g-adv-ntn-technology-for-\allowbreak
  direct-satellite-access-from-smart-phone-to-automotive-video-\allowbreak
  detailpage_251220-1597149.html}
\BIBentrySTDinterwordspacing

\bibitem{Li_Holographic_2025}
Q.~Li, M.~El-Hajjar, C.~Xu, K.~Li, X.~Feng, and L.~Hanzo, ``Holographic {MIMO}
  aided integrated user-centric cell-free terrestrial and non-terrestrial
  networks,'' \emph{IEEE Network}, pp. 1--1, 2025, early access.

\bibitem{Pan_Latency_2023_Sep}
G.~Pan, J.~Ye, J.~An, and M.-S. Alouini, ``Latency versus reliability in {LEO}
  {Mega}-constellations: Terrestrial, aerial, or space relay?'' \emph{IEEE
  Trans. Mob. Comput.}, vol.~22, no.~9, pp. 5330--5345, Sep. 2023.

\bibitem{Humphreys_Signal_2023_Oct}
T.~E. Humphreys, P.~A. Iannucci, Z.~M. Komodromos, and A.~M. Graff, ``Signal
  structure of the starlink {Ku}-band downlink,'' \emph{IEEE Trans. Aerosp.
  Electron. Syst.}, vol.~59, no.~5, pp. 6016--6030, Oct. 2023.

\bibitem{Huang_Airplane_2019_Sep}
X.~Huang, J.~A. Zhang, R.~P. Liu, Y.~J. Guo, and L.~Hanzo, ``Airplane-aided
  integrated networking for {6G} wireless: Will it work?'' \emph{IEEE Veh.
  Technol. Mag.}, vol.~14, no.~3, pp. 84--91, Sep. 2019.

\bibitem{Pan_Space_2023_Apr}
G.~Pan, H.~Zhang, R.~Zhang, S.~Wang, J.~An, and M.-S. Alouini, ``Space
  simultaneous information and power transfer: An enhanced technology for
  miniaturized satellite systems,'' \emph{IEEE Wireless Commun.}, vol.~30,
  no.~2, pp. 122--129, Apr. 2023.

\bibitem{Zhang_Space_2025_May}
C.~Zhang, Q.~Li, C.~Xu, L.-L. Yang, and L.~Hanzo, ``Space-air-ground integrated
  networks: Their channel model and performance analysis,'' \emph{IEEE Open J.
  Veh. Technol.}, vol.~6, pp. 1501--1523, May 2025.

\bibitem{3GPP_38.811}
``Study on {New Radio} ({NR}) to support non-terrestrial networks,'' {3GPP},
  Technical Report, TR 38.811 V15.4.0, Oct. 2020, release 15.

\bibitem{Patzold_modeling_1997_May}
M.~Patzold, U.~Killat, Y.~Li, and F.~Laue, ``Modeling, analysis, and simulation
  of nonfrequency-selective mobile radio channels with asymmetrical {Doppler}
  power spectral density shapes,'' \emph{{IEEE} Trans. Veh. Technol.}, vol.~46,
  no.~2, pp. 494--507, May 1997.

\bibitem{Liu_Multi-Scene_2021}
W.~Liu, S.~Zheng, Z.~Deng, K.~Wang, W.~Lin, J.~Lei, Y.~Jin, and H.~Liu,
  ``Multi-scene {Doppler} power spectrum modeling of {LEO} satellite channel
  based on atlas fingerprint method,'' \emph{IEEE Access}, vol.~9, pp.
  11\,811--11\,822, 2021.

\bibitem{Cheng_An_Adaptive_2009_Sep}
X.~Cheng, C.-X. Wang, D.~I. Laurenson, S.~Salous, and A.~V. Vasilakos, ``An
  adaptive geometry-based stochastic model for non-isotropic {MIMO}
  mobile-to-mobile channels,'' \emph{IEEE Trans. Wireless Commun.}, vol.~8,
  no.~9, pp. 4824--4835, Sep. 2009.

\bibitem{Ashby_Relativity_2003_Jan}
\BIBentryALTinterwordspacing
N.~Ashby, ``Relativity in the global positioning system,'' \emph{Living Reviews
  in Relativity}, vol.~6, no.~1, pp. 1--42, Jan. 2003. [Online]. Available:
  \url{https://link.springer.com/article/10.12942/lrr-2003-1}
\BIBentrySTDinterwordspacing

\bibitem{Cui_Multiobjective_2021_Sep}
J.~Cui, S.~X. Ng, D.~Liu, J.~Zhang, A.~Nallanathan, and L.~Hanzo,
  ``Multiobjective optimization for integrated ground-air-space networks:
  Current research and future challenges,'' \emph{IEEE Veh. Technol. Mag.},
  vol.~16, no.~3, pp. 88--98, Sep. 2021.

\bibitem{Yang_Doppler_2026}
\BIBentryALTinterwordspacing
L.-L. Yang, ``Doppler effect: Analyses and applications in wireless sensing and
  communications,'' \emph{arXiv preprint arXiv:2602.09955}, Feb. 2026.
  [Online]. Available: \url{https://arxiv.org/abs/2602.09955}
\BIBentrySTDinterwordspacing

\bibitem{Abdi_A_New_2003_May}
A.~Abdi, W.~Lau, M.-S. Alouini, and M.~Kaveh, ``A new simple model for land
  mobile satellite channels: first- and second-order statistics,'' \emph{IEEE
  Trans. Wireless Commun.}, vol.~2, no.~3, pp. 519--528, May 2003.

\bibitem{Tiejun_Performance_2026_Jun}
T.~Wang, J.~Proakis, E.~Masry, and J.~Zeidler, ``Performance degradation of
  {OFDM} systems due to {Doppler} spreading,'' \emph{IEEE Trans. Wireless
  Commun.}, vol.~5, no.~6, pp. 1422--1432, Jun. 2006.

\bibitem{Tang_Effect_1999_Dec}
X.~Tang, M.-S. Alouini, and A.~Goldsmith, ``Effect of channel estimation error
  on {M-QAM} {BER} performance in {Rayleigh} fading,'' \emph{IEEE Trans.
  Commun.}, vol.~47, no.~12, pp. 1856--1864, Dec. 1999.

\bibitem{Cao_Closed_2005_Jul}
L.~Cao and N.~Beaulieu, ``Closed-form {BER} results for {MRC} diversity with
  channel estimation errors in {Ricean} fading channels,'' \emph{IEEE Trans.
  Wireless Commun.}, vol.~4, no.~4, pp. 1440--1447, Jul. 2005.

\bibitem{Dong_Symbol_2005_Mar}
X.~Dong and L.~Xiao, ``Symbol error probability of two-dimensional signaling in
  {Ricean} fading with imperfect channel estimation,'' \emph{IEEE Trans. Veh.
  Technol.}, vol.~54, no.~2, pp. 538--549, Mar. 2005.

\bibitem{Bankey_Ergodic_2018_May}
V.~Bankey and P.~K. Upadhyay, ``Ergodic capacity of multiuser hybrid
  satellite-terrestrial fixed-gain {AF} relay networks with {CCI} and outdated
  {CSI},'' \emph{IEEE Trans. Veh. Technol.}, vol.~67, no.~5, pp. 4666--4671,
  May 2018.

\bibitem{Bankey_Performance_2018}
V.~Bankey, P.~K. Upadhyay, D.~B. Da~Costa, P.~S. Bithas, A.~G. Kanatas, and
  U.~S. Dias, ``Performance analysis of multi-antenna multiuser hybrid
  satellite-terrestrial relay systems for mobile services delivery,''
  \emph{IEEE Access}, vol.~6, pp. 24\,729--24\,745, 2018.

\bibitem{Liu_Frequency_2003_Mar}
Q.~Liu, ``Frequency synchronization in global satellite communications
  systems,'' \emph{IEEE Trans. Commun.}, vol.~51, no.~3, pp. 359--365, Mar.
  2003.

\bibitem{Zhou_A_Simultaneous_2025}
C.~Zhou, Z.~Tang, J.~Wei, and X.~Xia, ``A simultaneous positioning and orbit
  correction method using {Doppler} measurements of {LEO} satellites,''
  \emph{IEEE Trans. Instrum. Meas.}, vol.~74, pp. 1--19, 2025, early access.

\bibitem{Yeh_Efficient_2024_Dec}
B.-H. Yeh, J.-M. Wu, and R.~Y. Chang, ``Efficient {Doppler} compensation for
  {LEO} satellite downlink {OFDMA} systems,'' \emph{{IEEE} Trans. Veh.
  Technol.}, vol.~73, no.~12, pp. 18\,863--18\,877, Dec. 2024.

\bibitem{Tanash_Statistical_2025}
I.~M. Tanash, R.~Wichman, and N.~Gonz¨¢lez-Prelcic, ``Statistical modeling for
  accurate characterization of {Doppler} effect in {LEO}-terrestrial
  networks,'' \emph{IEEE Trans. Aerosp. Electron. Syst.}, pp. 1--14, 2025,
  early access.

\bibitem{Krieger_Relativistic_2014_Feb}
G.~Krieger and F.~De~Zan, ``Relativistic effects in bistatic synthetic aperture
  radar,'' \emph{IEEE Transactions on Geoscience and Remote Sensing}, vol.~52,
  no.~2, pp. 1480--1488, Feb. 2014.

\bibitem{Wang_GaussMask_2025}
S.~Wang, Z.~Song, Z.~Hua, X.~Yang, C.~Du, R.~Zhang, and G.~Pan,
  ``Gaussmask-{DSSS}: Enhancing covert spread spectrum communication with
  {Gaussian} cloaking and deep learning-aided synchronization,'' \emph{{IEEE}
  J. Sel. Areas Commun.}, pp. 1--1, 2025, early access.

\bibitem{Liu_Deep_2022_Apr}
D.~Liu, J.~Zhang, J.~Cui, S.-X. Ng, R.~G. Maunder, and L.~Hanzo, ``Deep
  learning aided routing for space-air-ground integrated networks relying on
  real satellite, flight, and shipping data,'' \emph{IEEE Wireless Commun.},
  vol.~29, no.~2, pp. 177--184, Apr. 2022.

\bibitem{Shi_Inverse_2017_May}
Z.~Shi, H.~Ding, S.~Ma, K.-W. Tam, and S.~Pan, ``Inverse moment matching based
  analysis of cooperative {HARQ-IR} over time-correlated {Nakagami} fading
  channels,'' \emph{IEEE Trans. Veh. Technol.}, vol.~66, no.~5, pp. 3812--3828,
  May 2017.

\bibitem{ITU-R_Refraction}
\BIBentryALTinterwordspacing
{International Telecommunication Union recommendation ITU-R}, ``The radio
  refractive index: its formula and refractivity data,'' pp. 453--459, Jul.
  2015. [Online]. Available:
  \url{https://www.itu.int/dms_pubrec/itu-r/rec/p/R-REC-P.453-11-201507-S!!PDF-E.pdf}
\BIBentrySTDinterwordspacing

\bibitem{EODG}
\BIBentryALTinterwordspacing
``Zenith absorption data,'' Earth Observation Data Group - University of
  Oxford, 2025, accessed: Mar. 13, 2025. [Online]. Available:
  \url{https://eodg.atm.ox.ac.uk/ATLAS/zenith-absorption}
\BIBentrySTDinterwordspacing

\bibitem{ROTHMAN_The_2013}
\BIBentryALTinterwordspacing
L.~Rothman, I.~Gordon, Y.~Babikov, A.~Barbe, D.~{Chris Benner}, P.~Bernath,
  M.~Birk, L.~Bizzocchi, V.~Boudon, L.~Brown, A.~Campargue, K.~Chance,
  E.~Cohen, L.~Coudert, V.~Devi, B.~Drouin, A.~Fayt, J.-M. Flaud, R.~Gamache,
  J.~Harrison, J.-M. Hartmann, C.~Hill, J.~Hodges, D.~Jacquemart, A.~Jolly,
  J.~Lamouroux, R.~{Le Roy}, G.~Li, D.~Long, O.~Lyulin, C.~Mackie, S.~Massie,
  S.~Mikhailenko, H.~Müller, O.~Naumenko, A.~Nikitin, J.~Orphal, V.~Perevalov,
  A.~Perrin, E.~Polovtseva, C.~Richard, M.~Smith, E.~Starikova, K.~Sung,
  S.~Tashkun, J.~Tennyson, G.~Toon, V.~Tyuterev, and G.~Wagner, ``The
  {HITRAN2012} molecular spectroscopic database,'' \emph{J. Quant. Spectrosc.
  Radiat. Transfer.}, vol. 130, pp. 4--50, 2013, hITRAN2012 special issue.
  [Online]. Available:
  \url{https://www.sciencedirect.com/science/article/pii/S0022407313002859}
\BIBentrySTDinterwordspacing

\bibitem{Curtis_OrbitalMechanics_2005}
H.~D. Curtis, \emph{Orbital Mechanics for Engineering Students}, 1st~ed.\hskip
  1em plus 0.5em minus 0.4em\relax Amsterdam, The Netherlands: Elsevier
  Butterworth-Heinemann, 2005.

\bibitem{AbramowitzStegun1965Handbook}
M.~Abramowitz and I.~A. Stegun, \emph{Handbook of Mathematical Functions with
  Formulas, Graphs, and Mathematical Tables}, ser. Applied Mathematics
  Series.\hskip 1em plus 0.5em minus 0.4em\relax Washington, DC, USA: U.S.
  Government Printing Office, 1965, vol.~55.

\bibitem{Ruiz_Geosynchronous_2013_Aug}
J.~Ruiz~Rodon, A.~Broquetas, A.~Monti~Guarnieri, and F.~Rocca, ``Geosynchronous
  {SAR} focusing with atmospheric phase screen retrieval and compensation,''
  \emph{IEEE Trans. Geosci. Remote Sens.}, vol.~51, no.~8, pp. 4397--4404, Aug.
  2013.

\bibitem{Ruiz_Nearly_2014_Oct}
J.~Ruiz-Rodon, A.~Broquetas, E.~Makhoul, A.~Monti~Guarnieri, and F.~Rocca,
  ``Nearly zero inclination geosynchronous {SAR} mission analysis with long
  integration time for earth observation,'' \emph{IEEE Trans. Geosci. Remote
  Sens.}, vol.~52, no.~10, pp. 6379--6391, Oct. 2014.

\bibitem{Diniz_An_Algorithm_2024_Oct}
A.~Diniz, T.~Eriksson, and U.~Gustavsson, ``An algorithm for harsh {Doppler}
  shift estimation for satellite communications,'' in \emph{Proc. Asilomar
  Conf. Signals, Systems, and Computers}, Oct. 2024, pp. 706--710.

\bibitem{Haas_Aeronautical_2002_Mar}
E.~Haas, ``Aeronautical channel modeling,'' \emph{{IEEE} Trans. Veh. Technol.},
  vol.~51, no.~2, pp. 254--264, Mar. 2002.

\bibitem{Hanzo_Adaptive_2002}
L.~Hanzo, C.~H. Wong, and M.~S. Yee, \emph{Adaptive Wireless Transceivers:
  Turbo-Coded, Turbo-Equalised and Space-Time Coded TDMA, CDMA and OFDM
  Systems}.\hskip 1em plus 0.5em minus 0.4em\relax Chichester, UK: John Wiley
  \& Sons, 2002.

\bibitem{Zhang_STAR_2022_Sep}
C.~Zhang, W.~Yi, Y.~Liu, Z.~Ding, and L.~Song, ``{STAR-IOS} aided {NOMA}
  networks: Channel model approximation and performance analysis,'' \emph{IEEE
  Trans. Wireless Commun.}, vol.~21, no.~9, pp. 6861--6876, Sep. 2022.

\bibitem{Continuous_Kotz_2000}
S.~Kotz, N.~Balakrishnan, and N.~L. Johnson, \emph{Continuous Multivariate
  Distributions: Models and Applications}, 2nd~ed., ser. Wiley Series in
  Probability and Statistics.\hskip 1em plus 0.5em minus 0.4em\relax New York,
  NY, USA: John Wiley \& Sons, Inc., 2000.

\bibitem{Press2007Numerical}
W.~H. Press, S.~A. Teukolsky, W.~T. Vetterling, and B.~P. Flannery,
  \emph{Numerical Recipes: The Art of Scientific Computing}, 3rd~ed.\hskip 1em
  plus 0.5em minus 0.4em\relax Cambridge, UK: Cambridge University Press, 2007.

\end{thebibliography}
}

\end{document}